\documentclass[%
 reprint,
nofootinbib,
 amsmath,amssymb,
 aps,
]{revtex4-2}

\usepackage{graphicx}
\usepackage{dcolumn}
\usepackage{bm}
\usepackage{braket}
\usepackage{hyperref}
\usepackage[dvipsnames]{xcolor}
\usepackage{slashed}

\newcommand{\Mpl}{{\rm M}_{\rm Pl}}

\newcommand{\Hquad}{\hspace{0.4em}} 
\newcommand{\as}{\slashed{a}}

\begin{document}

\preprint{APS/123-QED}

\title{(Re)constructing Accurate Axion Oscillations}

\author{Hoang Nhan Luu} \email{hoangnhan.luu@unh.edu}
\author{Chanda Prescod-Weinstein}
\affiliation{Department of Physics \& Astronomy, University of New Hampshire, Durham, NH 03824, USA}

\begin{abstract}

The cosmological evolution of ultralight axions typically involves rapid oscillations on the timescale of the inverse mass, making it challenging to resolve the dynamics at both the background and perturbation levels. In this study, we numerically implement a novel approach to this problem based on an effective field theory (EFT) -- developed by~\cite{Salehian:2020bon} and prepared for numerical implementation in~\cite{Luu:2026las} -- that is, for the first time, capable of reconstructing the relativistic oscillations intrinsic to the axion field. Compared to other available techniques, the reconstruction of the rapid oscillations is unique to the EFT approach, which describes the axion effective field through a wavefunction representation with relativistic corrections, rather than through effective fluid variables. From a computational standpoint, our approach archives a high level of accuracy -- up to a subpercent agreement with the exact solution -- while remaining relatively fast compared to alternative methods. As such, it opens up new possibilities for high-precision predictions for axion searches with future cosmological experiments.

\end{abstract}

\maketitle

\section{Introduction} \label{sec:introduction}

Despite the success of the standard model of particle physics, the nature of dark matter (DM) and dark energy (DE) remains among the greatest unsolved mysteries in modern physics. A plethora of evidence from the Cosmic Microwave Background (CMB)~\cite{Planck:2018vyg}, large-scale structures~\cite{Scognamiglio:2026phv} and astrophysical observations~\cite{SupernovaSearchTeam:1998fmf, Bradac:2008eu} suggests that these unknown components require new physics beyond the standard model~\cite{Abdalla:2022yfr}. Among the proposed hypotheses, axions stand out as one of the most motivated candidates for both DM and DE~\cite{Adams:2022pbo, Luu:2025fgw}.

In cosmology, ultralight axions behaving as a coherently oscillating scalar field are of particular interest~\cite{Hui:2021tkt, Ferreira:2020fam}, as their dynamics can be accurately described by classical equations of motion. The most widely studied example is the misalignment mechanism~\cite{Preskill:1982cy}, in which the axion field is initialized from a ``misaligned'' state (presumably set by inflation), and subsequently rolls down its potential to undergo damped oscillations around the vacuum. However, for most of the relevant mass range ($m \gg H_0$), these equations cannot be solved exactly all the way to redshift $z = 0$, as the relativistic axion field oscillates on the intrinsic timescale of $\mathcal{O}(m^{-1})$, much shorter than the Hubble timescale.

A common approach to this problem involves approximating axions as an effective fluid at late times~\cite{Ratra:1990me, Hwang:2009js}. This effective fluid then evolves according to its own system of effective equations that vary smoothly on the Hubble timescale. For instance, in the cosmological code \texttt{axionCAMB}~\cite{Hlozek:2014lca}, the effective fluid is characterized by the mean density, pressure, and velocity, obtained via \textit{cycle-averaging}. This procedure amounts to averaging the fluid quantities over one oscillation period, such that $\cos(mt) \rightarrow 0$ and $\cos^2(mt) \rightarrow 1/2$.

Broadly speaking, the idea is to solve the exact axion dynamics until some transition time, around the moment when the Hubble rate becomes comparable to the axion mass, $H \sim m$, and subsequently evolve the system within the effective regime until $z = 0$ or beyond. While \texttt{axionCAMB} was the first to adopt this procedure, the resulting evolution of the effective fluid turns out to depend on the choice of transition time between the exact and effective regimes~\cite{Cookmeyer:2019rna}. Such a dependence on the transition time implies a major drawback of this approach, as it is merely a precision parameter and should not influence the underlying physics.

Recently, a new approach to the effective fluid approximation (EFA) has been proposed by~\cite{Passaglia:2022bcr}, introducing several meaningful improvements that directly address the shortcomings of \texttt{axionCAMB}. First, and most notably, the matching conditions are carefully derived such that the effective fluid variables at the transition time strictly reflect their true time average. Second, the equation of state and sound speed of the effective fluid are calibrated from the analytical solution in the limit where the axion field is subdominant. As a result, one obtains axion equations of motion that depend on dynamical variables at the transition time. Ironically, these adjustments eliminate the dependence on the choice of transition time that was present in the previous approach. The full numerical implementation was later introduced in the cosmological code \texttt{AxiECAMB} by \cite{Liu:2024yne}.

Compared to \texttt{axionCAMB}, the evolution of the effective fluid is captured more accurately with \texttt{AxiECAMB}. However, the approach still has a minor caveat: the axion fluid variables and any quantities sourced by them, such as the Hubble function or metric perturbations, are discontinuous at the transition time. This is a natural and inevitable consequence, as one cannot ensure that oscillating variables and their time average coincide at all times. That said, discontinuities in axion fluid variables may cause unwanted implications, such as spurious Jeans oscillations~\cite{Liu:2024yne}, and should be avoided where possible.

Motivated by the limitations of existing approaches, we explore a novel strategy, based on the axion effective field theory (EFT) by \cite{Namjoo:2017nia, Salehian:2020bon, Salehian:2021khb}, to address the same problem. This work is the second in a series of two papers aimed at constructing a comprehensive framework for the EFT formalism. While the first paper~\cite{Luu:2026las} derived effective field equations for axions in the synchronous gauge, the current work is devoted to the numerical implementation of axion EFT in cosmological codes. Specifically, we will discuss in detail how this formalism can be incorporated into a realistic cosmological setting that includes other species alongside axions, such as those present in the standard $\Lambda$CDM model.

The rest of this paper is structured as follows. In Sec.~\ref{sec:theory}, we first review the effective theory of cosmological axions as presented in \cite{Luu:2026las}, focusing on the equations of motion for the so-called ``slow-mode wavefunction'' in \ref{sec:eft_axions}. We then present how the exact form of dynamical variables, such as axion density or metric perturbations, can be reconstructed from these slow modes in \ref{sec:reconstruction}, a feature unique to the EFT formalism. Section~\ref{sec:numerical_implementation} is dedicated to the main content of this study, namely the numerical implementation of axion EFT. It turns out that there are several non-trivial technicalities about this formalism that require special care in order to optimize computational speed while maintaining accurate results. These include the choice of initial conditions (\ref{sec:initial_conditions}), the transition time (\ref{sec:transition_time}), and the matching conditions (\ref{sec:matching_conditions}). Subtle problems related to the equations governing non-axion species (\ref{sec:non-axion_eqs}), as well as interpolation in common cosmological codes (\ref{sec:interpolation}) are also discussed here. In Sec.~\ref{sec:results}, we compare the results obtained from axion EFT with those of other approaches at both the background (\ref{sec:background_result}) and perturbation levels (\ref{sec:perturbations_result}). We also comment on the equivalence between the slow modes in EFT and the time-averaged quantities in these approaches (\ref{sec:slow-mode_result}). Finally, we conclude with a brief summary and outlook in Sec.~\ref{sec:conclusion}.

Our work adopt notations for the metric perturbations in the synchronous gauge from \cite{Ma:1995ey}. Natural units with $\hbar = c = 1$ and $\Mpl^2 = (8\pi G)^{-1}$ are implicitly assumed everywhere.

\section{Theory} \label{sec:theory}

Since the theoretical background of the effective field theory for axions has been detailed in \cite{Luu:2026las, Salehian:2020bon}, we only summarize the results relevant for numerical implementation. For interested readers, the derivation of all equations in this section can be found in our companion paper~\cite{Luu:2026las}. Note that, unlike here, they are presented in that work in terms of rescaled dimensionless variables for ease of computation. Specifically, to convert the results from \cite{Luu:2026las} to those in this section, we use
\begin{align}
\begin{gathered}
    \tilde{t} = mt, \quad \tilde{H} = \dfrac{H}{m}, \quad \tilde{\psi} = \dfrac{\bar{\psi}}{\sqrt{m}\Mpl}, \\ \delta\tilde{\psi} = \dfrac{\delta\psi}{\sqrt{m}\Mpl}, \quad \tilde{\rho} = \dfrac{\bar{\rho}}{m^2\Mpl^2}, \quad \tilde{p} = \dfrac{\bar{p}}{m^2\Mpl^2}, \\
    \delta\tilde{\rho} = \dfrac{\delta\rho}{m^2\Mpl^2}, \quad \delta\tilde{p} = \dfrac{\delta p}{m^2\Mpl^2}, \quad \delta\tilde{U} = \dfrac{\delta U}{m\Mpl^2},
\end{gathered} \label{eq:tilde_variables}
\end{align}
with the rescaled time derivative $X' = \partial X/\partial\tilde{t} = m^{-1}\dot{X}$.

\subsection{Effective Field Theory for axions} \label{sec:eft_axions}

In the most minimal model without non-trivial couplings, the cosmological axion field can be described by a scalar field $\phi$ of mass $m$. The dynamical time scales of such a field is of order $m^{-1}$, which motivates a change of variable from the real field $\phi$ to the complex \textit{wavefunction} $\psi$~\cite{Salehian:2020bon}, satisfying
\begin{align}
    &\phi = \dfrac{1}{\sqrt{2m}} \left( \psi e^{-imt} + \psi^*e^{imt} \right), \label{eq:phi} \\
    &\dot{\phi} = -i\sqrt{\dfrac{m}{2}} \left( \psi e^{-imt} - \psi^* e^{imt} \right), \label{eq:phi_dot}
\end{align}
where ``dot'' denotes derivatives w.r.t the proper time $t$. As we split $\psi(t,\bm{x}) = \bar{\psi}(t) + \delta\psi(t,\bm{x})$, the background equation of motion is given by
\begin{align}
    \dot{\bar{\psi}} = - \dfrac{3}{2}H\bar{\psi} + \dfrac{3}{2}H\bar{\psi}^*e^{2imt}, \label{eq:psi_exact_background}
\end{align}
and the perturbation equation at the linear regime is
\begin{multline}
    \dot{\delta\psi} = - \left( \dfrac{3}{2}H + \dfrac{ik^2}{2ma^2} \right)\delta\psi + \dfrac{1}{4}\bar{\psi}\dot{h} \\
    + e^{2imt}\left[ \left( \dfrac{3}{2}H - \dfrac{ik^2}{2ma^2} \right)\delta\psi^* + \dfrac{1}{4}\bar{\psi}^*\dot{h} \right]. \label{eq:dpsi_exact_perturbations}
\end{multline}
The latter equation is derived in the momentum space of the synchronous gauge. We use the convention $h$ and $\eta$ to denote metric perturbations as in \cite{Ma:1995ey} and $H$ is the Hubble function.

It is obvious that equations \eqref{eq:psi_exact_background} and \eqref{eq:dpsi_exact_perturbations} are oscillatory by their dependence on the explicit exponential factors $e^{2imt}$. Therefore, solving these equations requires special treatment when axion oscillations become faster than the Hubble time scale of order $H^{-1}$. The effective field theory (EFT) method suggests expanding every dynamical variables with a mode expansion
\begin{align}
    X(t,\bm{x}) = \sum^{\infty}_{\nu=-\infty} X_\nu(t,\bm{x}) e^{i\nu mt}, \label{eq:nu_expansion}
\end{align}
where the $\nu = 0$ term corresponds to the \textit{slow mode} that vary on the Hubble time scale instead of $m^{-1}$. As such, the equations of motion for $X_0 \equiv X_s$ are non-stiff and trivial to solve numerically. In the case of the slow-mode wavefunction $\psi_s$, we have derived the background equation~\cite{Salehian:2020bon, Luu:2026las}
\begin{multline}
    \dot{\bar{\psi}}_s = -\dfrac{3}{2}H_s\bar{\psi}_s + \dfrac{3i}{16m\Mpl^2}\bar{\psi}_s \left( 3m|\bar{\psi}_s|^2 + 2\bar{\rho}_{\as,s} \right) \\
    - \dfrac{9}{32m^2\Mpl^2}H_s\bar{\psi}_s \left( m|\bar{\psi}_s|^2 + \bar{\rho}_{\as,s} + \bar{p}_{\as,s} \right) \label{eq:psi_slow}
\end{multline}
and perturbation equation in the synchronous gauge~\cite{Luu:2026las}
\begin{multline}
    \dot{\delta\psi}_s = - \left( \dfrac{3}{2}H_s + \dfrac{ik^2}{2ma^2_s} \right)\delta\psi_s - \dfrac{1}{4}\dot{h}_s\bar{\psi}_s \\ + \left( \dfrac{3iH_s}{8m} + \dfrac{k^2}{16m^2a_s^2} \right)\bar{\psi}_s \dot{h}_s + \dfrac{3i\bar{\psi}_s^{*2}}{16\Mpl^2}\delta\psi_s^* \\
    + \left( \dfrac{9iH_s^2}{8m} + \dfrac{3i|\bar{\psi}_s|^2}{8\Mpl^2} + \dfrac{ik^4}{8m^3a_s^4} \right)\delta\psi_s. \label{eq:dpsi_slow}
\end{multline}
Here, we use the subscript ``$a$'' to indicate axion-related variables and ``$\as$'' for the collective sum of common species other than the axion field.\footnote{They include: photons, baryons, neutrinos, cold dark matter and cosmological constant in the $\Lambda$CDM model.} In order to supplement the above system, we also need an EFT version for the Hubble function and metric perturbations. The first Friedmann equation yields
\begin{align}
    3\Mpl^2H_s^2 = m|\bar{\psi}_s|^2 + \bar{\rho}_{\as,s} + \dfrac{3|\bar{\psi}_s^2|}{32m\Mpl^2}\left( m|\bar{\psi}_s|^2 + 2\bar{\rho}_{\as,s} \right). \label{eq:H_slow}
\end{align}
From $00$ and $0i$ Einstein equations, we then have
\begin{align}
    \Mpl^2 \dot{h}_s &= \dfrac{2\Mpl^2k^2}{a_s^2H_s}\eta_s + \dfrac{m}{H_s}\left(\bar{\psi}^*_s\delta\psi_s + \bar{\psi}_s\delta\psi_s^*\right) + \dfrac{\delta\rho_{\as,s}}{H_s}, \label{eq:hdot_slow} \\
    \Mpl^2\dot{\eta}_s &= \dfrac{im}{4}\left(\bar{\psi}^*_s\delta\psi_s - \bar{\psi}_s\delta\psi_s^*\right) - \dfrac{m}{2}\delta U_{\as,s} \nonumber \\
    &\hspace{0.8cm} - \dfrac{1}{16}|\bar{\psi}_s|^2\dot{h}_s - \dfrac{3}{8}H_s\left(\bar{\psi}^*_s\delta\psi_s + \bar{\psi}_s\delta\psi_s^*\right) \nonumber \\
    &\hspace{1.2cm} - \dfrac{ik^2}{16ma_s^2}\left(\bar{\psi}^*_s\delta\psi_s - \bar{\psi}_s\delta\psi_s^*\right). \label{eq:eta_slow}
\end{align}

\subsection{Reconstruction of exact oscillatory features} \label{sec:reconstruction}

In axion EFT, the slow modes of dynamical variables are equivalent to the time-averaged or effective quantities in other approaches~\cite{Hlozek:2014lca, Passaglia:2022bcr}. By definition
\begin{align}
    X_s(t,\bm{x}) \equiv \dfrac{m}{2\pi}\int_{-\infty}^{\infty} {\rm sinc}\left[\dfrac{m}{2}(t-t')\right] X(t,\bm{x})dt',
\end{align}
where the sinc function serves as a weighted factor. Thus, $X_s$ is not exactly the cycle-averaging of $X$ because it also includes small (but non-negligible) contributions from the two tails of the sinc function. In the frequency space, these contributions are equivalent to the \textit{backreactions} of higher-frequency modes $X_\nu$ with $\nu \neq 0$ contained within $-m/2$ and $m/2$~\cite{Salehian:2020bon}.

The \textit{relativistic corrections} are then defined as these backreactions, expressed in terms of some ``small'' parameters, collectively denoted as $\epsilon \ll 1$. The two most important examples of such parameters are
\begin{align}
    \epsilon_H = \dfrac{H_s}{m} \quad \text{and} \quad \epsilon_k = \dfrac{k^2}{m^2a_s^2}. \label{eq:transition_condition}
\end{align}
Other examples are usually expressed as the dimensionless rescaled version of the slow-mode variables (which are related to those described in \eqref{eq:tilde_variables}) such as
\begin{align}
    \dfrac{\bar{\psi}_s}{\sqrt{m}\Mpl}, \quad \dfrac{\delta\psi_s} {\sqrt{m}\Mpl}, \quad \eta, \quad \dfrac{\dot{h}}{m}. \label{eq:transition_condition_2}
\end{align}
These quantities are considered ``first order'' in $\epsilon$, {\it i.e.}, they are of the same order of relativistic corrections with $\epsilon_H$ and $\epsilon_k$ above. Applying perturbation-series approximation in the small parameter $\epsilon$, we can compute relativistic corrections of any dynamical variables up to arbitrarily high orders~\cite{Luu:2026las}. 

As a consequence, it is straightforward to \textit{reconstruct} the exact quantities from their slow modes with the expansion \eqref{eq:nu_expansion}. Explicitly, the exact wavefunction can be reconstructed as
\begin{multline}
    \bar{\psi} = \bar{\psi}_s - \dfrac{3iH_s}{4m} \bar{\psi}^*_s e^{2imt} \\ - \dfrac{3\bar{\psi}^*_s}{32m^2\Mpl^2}\left[ m|\bar{\psi}_s|^2 + 2 \left( \bar{\rho}_{\as,s} + \bar{p}_{\as,s} \right) \right]e^{2imt} \\ + \dfrac{3\bar{\psi}^3}{32m\Mpl^2}e^{-2imt} - \dfrac{3\bar{\psi}_s^{*3}}{64m\Mpl^2}e^{4imt} \label{eq:reconstructed_psi}
\end{multline}
at the background level and
\begin{align}
\delta\psi = \delta\psi_s - \left[ \left( \dfrac{3iH_s}{4m} + \dfrac{k^2}{4m^2a_s^2} \right)\delta\psi^*_s + \dfrac{i\dot{h}_s}{8m}\bar{\psi}_s^* \right]e^{2imt} \label{eq:reconstructed_dpsi}
\end{align}
at the perturbation level. Compared to the leading order of $\mathcal{O}(\epsilon)$, we have included terms up to two higher-order corrections, {\it i.e.}, $\mathcal{O}(\epsilon^3)$, at the background level but only up to one higher-order corrections, {\it i.e.}, $\mathcal{O}(\epsilon^2)$, at the perturbation level. This is why the background wavefunction \eqref{eq:reconstructed_psi} looks more complicated than \eqref{eq:reconstructed_dpsi} for its perturbations. The rationale behind this choice is two-fold. On the one hand, since corrections at next-to-next-leading order, namely $\mathcal{O}(\epsilon^3)$ or higher in this case, are very small within the applicable range of the EFT, it is usually unnecessary to include anything beyond next-to-leading order corrections; this is what we adopt for perturbations in this work. On the other hand, since nearly all perturbation equations require knowledge of the background variables, any errors in these quantities would be amplified in the reconstruction of perturbations, making it particularly beneficial to include additional corrections for the background variables.

One can also reconstruct the exact Hubble function
\begin{align}
    H = H_s - \dfrac{i}{8\Mpl^2} \left( \bar{\psi}_s^{*2}e^{2imt} - \bar{\psi}_s^2 e^{-2imt} \right)
\end{align}
as well as the exact metric perturbations
\begin{align}
    &\dot{h} = \dot{h}_s - \dfrac{3i}{2\Mpl^2} \left( \bar{\psi}^*_s\delta\psi_s^*e^{2imt} - \bar{\psi}_s\delta\psi_se^{-2imt} \right), \label{eq:reconstructed_hdot} \\
    &\eta = \eta_s + \dfrac{1}{8m\Mpl^2} \left( \bar{\psi}^*_s\delta\psi_s^*e^{2imt} + \bar{\psi}_s\delta\psi_se^{-2imt} \right). \label{eq:reconstructed_eta}
\end{align}

The slow modes of some variables can be approximated by their exact counterparts because the first finite corrections only come in at the next-to-next-leading order~\cite{Luu:2026las}. These include the scale factor and the fluid variables of non-axion species
\begin{equation}
    \begin{gathered}
        a_s \sim a, \quad \bar{\rho}_{\as,s} \sim \bar{\rho}_{\as}, \quad \bar{p}_{\as,s} \sim \bar{p}_{\as} \\
        \delta\rho_{\as,s} \sim \delta\rho_{\as}, \quad \delta p_{\as,s} \sim \delta p_{\as}, \quad \delta U_{\as,s} \sim \delta U_{\as} \\
        [(\bar{\rho}_{\as} + \bar{p}_{\as})\sigma_{\as}]_s \sim (\bar{\rho}_{\as} + \bar{p}_{\as})\sigma_{\as},
    \end{gathered} \label{eq:nonaxion_variables}
\end{equation}
where $\sigma_{\as}$ in the last row denotes the anisotropic stress (largely contributed by photons and neutrinos).

Having the wavefunction reconstructed, the exact fluid description of the axion field can be easily computed as in any other methods. Firstly, the axion background density and pressure are given by
\begin{align}
    &\bar{\rho}_a = m|\bar{\psi}|^2, \label{eq:rho_exact} \\
    &\bar{p}_a = - \dfrac{m}{2}\left( \bar{\psi}^2 e^{-2imt} + \bar{\psi}^{*2}e^{2imt} \right). \label{eq:p_exact}
\end{align}
Similarly, we have the axion density, pressure, and velocity perturbations in the synchronous gauge as follows
\begin{align}
    &\delta\rho_a = m \left( \bar{\psi}^*\delta\psi + \bar{\psi}\delta\psi^* \right), \\
    &\delta p_a = -m \left( \bar{\psi}\delta\psi e^{-2imt} + \bar{\psi}^*\delta\psi^* e^{2imt} \right), \\
    &\delta U_a = \dfrac{i}{2}\left( \bar{\psi}\delta\psi^* - \bar{\psi}^*\delta\psi + \bar{\psi}\delta\psi e^{-2imt} - \bar{\psi}^*\delta\psi^*e^{2imt} \right),
\end{align}
where $\delta U_a \equiv - a(\bar{\rho}_a + \bar{p}_a)v_a/k$. As we can see, the fluid variables are completely determined once $\bar{\psi}$ and $\delta\psi$ are known, so there is no need to reconstruct them separately. Optionally, one may also want to know the associated slow modes of these variables for use in non-axion equations of motion (see our discussion in Sec.~\ref{sec:non-axion_eqs}). They are given by~\cite{Luu:2026las}
\begin{align}
    &\bar{\rho}_{a,s} = m|\bar{\psi}_s|^2 + \dfrac{3}{16m\Mpl^2}\left( m|\bar{\psi}_s|^2 + \bar{\rho}_{\as,s} \right)|\bar{\psi}_s|^2, \\
    &\bar{p}_{a,s} = \dfrac{3}{16m\Mpl^2}\left( m|\bar{\psi}_s|^2 + 2\bar{\rho}_{\as,s} + 2\bar{p}_{\as,s} \right)|\bar{\psi}_s|^2, \\
    &\delta\rho_{a,s} = m\left( \bar{\psi}_s^*\delta\psi_s + \bar{\psi}_s\delta\psi_s^* \right),  \\
    &\delta p_{a,s} = \dfrac{k^2}{4ma_s^2}\left( \bar{\psi}_s^*\delta\psi_s + \bar{\psi}_s\delta\psi_s^* \right), \\
    &\delta U_{a,s} = - \dfrac{i}{2}\left( \bar{\psi}_s^*\delta\psi_s - \bar{\psi}_s\delta\psi_s^* \right) + \dfrac{3H_s}{4m}\left( \bar{\psi}_s^*\delta\psi_s + \bar{\psi}_s\delta\psi_s^* \right) \nonumber \\
    &\hspace{1.2cm} + \dfrac{ik^2}{8m^2a_s^2}\left( \bar{\psi}_s^*\delta\psi_s - \bar{\psi}_s\delta\psi_s^* \right) + \dfrac{1}{8m}|\bar{\psi}_s|^2\dot{h}_s.
\end{align}

\section{Numerical implementation} \label{sec:numerical_implementation}

The EFT formalism presented in the previous sections could be integrated in common Boltzmann solvers, such as \texttt{CAMB}~\cite{Howlett:2012mh} or \texttt{CLASS}~\cite{Blas:2011rf}. In this section we discuss how to implement the EFT formalism in such cosmological codes. As a demonstration, we incorporate the implementation presented here into a new package called \texttt{CLAxions}\footnote{At the time of wrting this paper, \texttt{CLAxions} is not yet publicly available but will be made available at a future date.}, namely ``CLASS for Axions.'' Conceptually, the algorithm proceeds through the following steps:
\begin{itemize}
    \item \textbf{Step 1}: Choose an appropriate initial time $t_i$ and a transition time $t_*$. We may loosely refer to $a_i$ and $a_*$ (scale factor), or $\tau_i$ and $\tau_*$ (conformal time) as ``time'' from now on.
    \item \textbf{Step 2}: Evolve the exact equations of motion from $t_i$ to $t_*$. These include \eqref{eq:psi_exact_background}, \eqref{eq:dpsi_exact_perturbations} for the wavefunction and Einstein equations governing $\eta$ and $\dot{h}$.
    \item \textbf{Step 3}: Find the matching conditions for the slow modes at $t_*$.
    \item \textbf{Step 4}: Evolve the EFT equations from $t_*$ till the present day. These include \eqref{eq:psi_slow}, \eqref{eq:dpsi_slow} for the slow-mode wavefunction and \eqref{eq:eta_slow}, \eqref{eq:hdot_slow} the slow-mode metric perturbations.
    \item \textbf{Step 5}: Reconstruct fluid variables and compute cosmological observables.
\end{itemize}
Except for the reconstruction part, these steps are quite standard among other cosmological codes for axions, such as \texttt{axionCAMB}~\cite{Hlozek:2014lca} or more recently \texttt{AxiECAMB}~\cite{Liu:2024yne}. There are, however, several subtleties that merit careful attention when implementing the EFT approach in a realistic cosmological setting. We discuss them one-by-one below.

\subsection{Initial conditions} \label{sec:initial_conditions}

Since the EFT equations are only valid in the non-relativistic limit when certain parameters are small, especially those in \eqref{eq:transition_condition}, we cannot solve them  when the radiation density still dominates the Hubble flow, which makes $\epsilon_H \gg 1$ and possibly $\epsilon_k \gg 1$ for certain wavenumbers. Luckily, the exact equations are still solvable in this regime given the transition time is not too late. In the simplest case where axion fluctuations are assumed to be adiabatic, their initial conditions can be computed with the power-series method (see App.~\ref{app:power_series_ICs})
\begin{align}
    &\Omega_a \rightarrow \bar{\phi}_i, \quad \quad \dot{\bar{\phi}}_i = - \dfrac{m^2\mathcal{C}^2}{5a}\bar{\phi}_i\tau_i^3, \label{eq:ics_background} \\
    &\delta\phi_i = \dfrac{m^2k^2\mathcal{C}^2}{420}\bar{\phi}_i\tau_i^6, \quad \dot{\delta\phi}_i = \dfrac{m^2k^2\mathcal{C}^2}{70a}\bar{\phi}_i\tau^5_i, \label{eq:ics_perturbations}
\end{align}
where $\bar{\phi}_i$ is the initial field value, $\tau$ is the conformal time, $\mathcal{C} = H_0\sqrt{\Omega_r}$, assuming that the axion field is initialized in radiation domination era and $\Omega_r$ is the fractional radiation density at the present.

There are a few remarks regarding the implementation of these initial conditions in practice. First, $\bar{\phi}_i$ is not known \textit{a priori} because it determines the specific amount of axions in the total matter budget. In \texttt{CLAXions}, we use the shooting method to infer $\bar{\phi}_i$ from the present-day fractional axion density $\Omega_a$, so that $\Omega_a$ is one of the input parameters (beside the axion mass $m$). Second, the power-series solutions apply only when $\tau_i$ is chosen before the background field $\bar{\phi}$ becomes dynamical (which is characterized by the transition time). In other words, all $k$-modes of axion perturbations must be initialized before $t_*$ for Eqs.~\eqref{eq:ics_background} and \eqref{eq:ics_perturbations} to be valid. Third, although the initial conditions are given in terms of $\phi$'s and its derivatives, the initial wavefunction can be inferred from \eqref{eq:phi} and \eqref{eq:phi_dot}. Specifically, we have
\begin{align}
    {\rm Re}(\psi_i) &= \sqrt{\dfrac{m}{2}} \cos(mt_i)\phi_i - \dfrac{1}{\sqrt{2m}}\sin(mt_i)\dot{\phi}_i, \\
    {\rm Im}(\psi_i) &= \sqrt{\dfrac{m}{2}} \sin(mt_i)\phi_i + \dfrac{1}{\sqrt{2m}}\cos(mt_i)\dot{\phi}_i
\end{align}
for the real and imaginary part of $\psi_i$. Thus, $\bar{\psi}_i$ and $\delta\psi_i$ are obtained by replacing $\phi_i$ with $\bar{\phi}_i$ and $\delta\phi_i$, respectively. Using ${\rm Re}(\psi)$ and ${\rm Im}(\psi)$ is more convenient for numerical codes that do not support complex numbers like \texttt{CLASS} does, but $\psi$ and $\psi^*$ is another choice to specify wavefunction degrees of freedom.

\subsection{Transition time} \label{sec:transition_time}

The transition time $t_*$ should be chosen when $\epsilon_H$ and $\epsilon_k$ in \eqref{eq:transition_condition} are considered ``small'', {\it i.e.}, at least less than $1$ (smaller is better). This is a single most important requirement for the validity of the axion EFT. Background evolution only requires $\epsilon_H$ condition, but perturbations evolution would need both satisfied. As such, let us define the transition time $t^{b}_*$ such that $\epsilon_H(t^b_*) = E_H$ for the background, and $t^p_* = {\rm max}(t_1,t_2)$ where $\epsilon_H(t_1) = E_H$ and $\epsilon_k(t_2) = E_k$ for the perturbations. Here, $E_H$ and $E_k$ are called ``precision'' parameters\footnote{A class of parameters controlling the accuracy of numerical computations (but not the underlying physics) in \texttt{CLASS}}. By default, \texttt{CLAxions} sets both parameters to 0.1 to balance speed and accuracy, though they can be adjusted to higher or lower values as needed. Despite this freedom, $t^b_*$ and $t^p_*$ might not be the same, especially for perturbation modes with high $k$ and small $m$.

For instance, consider an ultralight axion field with $m = 10^{-25}~{\rm eV}$, the condition $\epsilon_H(t^b_*) = H_*/m = 0.1$ gives the scale factor at the transition time $a^b_* \sim 3.8 \times 10^{-5}$. Plugging $a^b_*$ into \eqref{eq:transition_condition}, we find $\epsilon_k(t^b_*) = (k/ma^b_*)^2 \sim 2.8$ for $k = 1~{\rm Mpc}^{-1}$. As $\epsilon_k(t^b_*) > 1$, this time could not be chosen as the transition time for the perturbations. Otherwise, the last term in equation \eqref{eq:dpsi_slow}, $k^4/m^4a_s^4$ after rescaling, is no longer considered ``small'' compared to others. Now, computing the scale factor at $\epsilon(t^p_*) = 0.1$ we obtain $a^p_* \sim 2 \times 10^{-4}$, implying a much later time than the background transition. From this example, it is obvious to see that each $k$ mode with $t^p_* > t^b_*$ has a different transition time. Although we can choose $t^b_*$ equal to $t^p_*$ of the largest $k$ of our interest so that there is only one transition time, this would drastically slow down the computations when all smaller $k$-modes need to evolve with stiff equations for much longer. The EFT approach is, therefore, slower than other methods where the second condition with $\epsilon_k$ is not required. This additional computational cost is compensated by the greater quantitative precision of the EFT solutions, as we will show in Sec.~\ref{sec:slow-mode_result}.

\subsection{Matching conditions} \label{sec:matching_conditions}

The EFT formalism offers a unique way to smoothly match dynamical variables between two regimes. Since the exact variables could be reconstructed from their slow modes, the idea is to equate the exact wavefunction to the reconstructed one at the transition time. That means, by inverting Eqs.~\eqref{eq:reconstructed_psi} and \eqref{eq:reconstructed_dpsi}, we can find the values of $\bar{\psi}_s(t^b_*)$ and $\delta\psi_s(t^p_*)$, given $\bar{\psi}(t^b_*)$ and $\delta\psi(t^p_*)$ already known from the exact solutions.

The inverting problem may not be trivial because we need to solve a non-linear system of equations when higher-order corrections are involved. Firstly, let us consider Eq.~\eqref{eq:reconstructed_psi} for the background wavefunction. Decomposing this equation into its real and imaginary part yields
\begin{align}
    &(1 + A_{\rm sin} )\Re_s - A_{\rm cos} \Im_s = \Re + C_{\rm re}, \label{eq:matching_psi_1} \\
    &(1 - A_{\rm sin})\Im_s - A_{\rm cos}\Re_s = \Im + C_{\rm im}, \label{eq:matching_psi_2}
\end{align}
To avoid cluttered notation, we have introduced the following shorthand:
\begin{align}
\begin{gathered}
    \Re = {\rm Re}(\bar{\psi}), \quad \Im = {\rm Im}(\bar{\psi}), \\
    \Re_s = {\rm Re}(\bar{\psi}_s), \quad \Im_s = {\rm Im}(\bar{\psi}_s), \\
    c_{jm} = \cos(jmt), \quad s_{jm} = \sin(jmt).
\end{gathered}
\end{align}
Additionally, $A_{\rm sin}$ and $A_{\rm cos}$ are given by
\begin{align}
    A_{\rm sin} = \dfrac{3H_s}{4m}s_{2m}, \quad A_{\rm cos} = \dfrac{3H_s}{4m}c_{2m},
\end{align}
whereas $C_{\rm re}$ and $C_{\rm im}$ include a non-linear combinations of terms in the second and third row of \eqref{eq:reconstructed_psi}
\begin{align}
    &C_{\rm re} = \dfrac{3}{64m\Mpl^2}  \left[c_{4m}\Re^3_s + (4s_{2m} - s_{4m})\Im^3_s \right. \nonumber \\
    &\hspace{1.2cm} \left. - (4s_{2m} - 3s_{4m})\Re^2_s\Im_s + (8c_{2m} - 3c_{4m})\Re_s\Im^2_s \right] \nonumber \\
    &\hspace{1.5cm} + \dfrac{3}{16m^2\Mpl^2} (\bar{\rho}_{\as,s} + \bar{p}_{\as,s})(c_{2m}\Re_s + s_{2m}\Im_s) \label{eq:C_re} \\
    &C_{\rm im} = \dfrac{3}{64m\Mpl^2}  \left[(4s_{2m} + s_{4m})\Re^3_s + c_{4m}\Im^3_s \right. \nonumber \\
    &\hspace{1.2cm} \left. - (8c_{2m} + 3c_{4m})\Re^2_s\Im_s - (4s_{2m} + 3s_{4m})\Re_s\Im^2_s \right] \nonumber \\
    &\hspace{1.5cm} + \dfrac{3}{16m^2\Mpl^2} (\bar{\rho}_{\as,s} + \bar{p}_{\as,s})(s_{2m}\Re_s - c_{2m}\Im_s) \label{eq:C_im}
\end{align}
If $A$'s and $C$'s can be treated as constants, \eqref{eq:matching_psi_1} and \eqref{eq:matching_psi_2} would form a linear algebraic system of equations with the solution given by
\begin{align}
    \begin{pmatrix}
        \Re_s \\ \Im_s 
    \end{pmatrix} = 
    \begin{pmatrix}
        1 + A_{\rm sin} & -A_{\rm cos} \\ - A_{\rm cos} & 1 - A_{\rm sin} 
    \end{pmatrix}^{\!\!-1}
    \begin{pmatrix}
        \Re + C_{\rm re} \\ \Im + C_{\rm im}
    \end{pmatrix}. \label{eq:matching_re_im_psi}
\end{align}
In fact, by approximating non-constant terms in $A$'s and $C$'s by their exact counterparts at $t = t^b_*$
\begin{equation}
    \begin{gathered}
        \Re_s(t^b_*) \sim \Re(t^b_*), \Hquad \Im_s(t^b_*) \sim \Im(t^b_*), \Hquad H_s(t_*^b) \sim H(t_*^b),
    \end{gathered}
\end{equation}
the solution for $\Re_s(t^b_*), \Im_s(t^b_*)$ obtained with \eqref{eq:matching_re_im_psi} is not far off from the correct one. In order to refine this solution, we repeat this procedure with the updated value of $\Re_s, \Im_s$, {\it i.e.}, substituting the $\Re_s, \Im_s$ values just found back to \eqref{eq:H_slow}, \eqref{eq:C_re}, \eqref{eq:C_re} to find the new $H_s, C_{\rm re}, C_{\rm im}$ and apply Eq.~\eqref{eq:matching_re_im_psi} again until they converge. Since non-linear terms, such as $\Re^3_s$ or $\Im^3_s$, already become quite small at the transition time, we only need to iterate 2-3 times to obtain the solution with a relative error of order $\sim \mathcal{O}(10^{-8})$\footnote{This is numerical error comes from solving \eqref{eq:matching_psi_1} and \eqref{eq:matching_psi_2}. It is not the error between $\bar{\psi}_s$ found with those equations and $\bar{\psi}_s$ as the exact average of $\bar{\psi}$, which depends on how many higher-order corrections are included in \eqref{eq:reconstructed_psi} (see \cite{Luu:2026las} for more details).}.

The matching conditions for perturbations can be derived in a similar manner. From \eqref{eq:reconstructed_dpsi} we have
\begin{align}
    &(1 + B_- )\delta\Re_s - B_+\delta\Im_s - D_-\dot{h}_s = \delta\Re, \label{eq:matching_dpsi_1} \\
    &(1 - B_-)\delta\Im_s - B_+\delta\Re_s - D_+\dot{h}_s = \delta\Im, \label{eq:matching_dpsi_2}
\end{align}
where the notations follow as above, {\it i.e.},
\begin{align}
\begin{gathered}
    \delta\Re = {\rm Re}(\delta\psi), \quad \delta\Im = {\rm Im}(\delta\psi), \\
    \delta\Re_s = {\rm Re}(\delta\psi_s), \quad \delta\Im_s = {\rm Im}(\delta\psi_s).
\end{gathered}
\end{align}
$B$'s and $D$'s variables are given as
\begin{align}
    B_{\rm +} &= \dfrac{3H_s}{4m}c_{2m} + \dfrac{k^2}{4m^2a_s^2}s_{2m}, \\
    B_{\rm -} &= \dfrac{3H_s}{4m}s_{2m} - \dfrac{k^2}{4m^2a_s^2}c_{2m}, \\
    D_{\rm +} &= \dfrac{1}{8m}\left( c_{2m}\Re_s + s_{2m}\Im_s \right), \\
    D_{\rm -} &= \dfrac{1}{8m}\left( c_{2m}\Im_s - s_{2m}\Re_s \right).
\end{align}
Notice that $\dot{h}_s$ should be solved besides $\delta\Re_s, \delta\Im_s$ for perturbations. Thus, we derive another condition from \eqref{eq:reconstructed_hdot}, which reads
\begin{align}
    \dot{h}_s - E_-\delta\Re_s - E_+\delta\Im_s =  \dot{h}, \label{eq:matching_hdot}
\end{align}
where $E$'s variables are
\begin{align}
    E_+ &= \dfrac{3}{\Mpl^2}\left( \Im_s s_{2m} + \Re_s c_{2m} \right), \\
    E_- &= \dfrac{3}{\Mpl^2}\left( \Im_s c_{2m} - \Re_s s_{2m} \right).
\end{align}
Eqs.~\eqref{eq:matching_dpsi_1}, \eqref{eq:matching_dpsi_2}, \eqref{eq:matching_hdot} together admit the solution
\begin{align}
    \begin{pmatrix}
        \delta\Re_s \\ \delta\Im_s \\ \dot{h}_s
    \end{pmatrix} = 
    \begin{pmatrix}
        1 + B_- & - B_+ &-D_- \\
        - B_+ & 1 - B_- & - D_+ \\
        - E_- & - E_+ & 1
    \end{pmatrix}^{\!\!-1}
    \begin{pmatrix}
        \delta\Re \\ \delta\Im \\ \dot{h}
    \end{pmatrix}. \label{eq:matching_re_im_dpsi}
\end{align}
Since $B$'s, $D$'s and $E$'s only depend on background quantities computed previously, there is no need for iteration. However, that technique still comes in handy in case we would like to include higher-order corrections in \eqref{eq:reconstructed_dpsi}, \eqref{eq:reconstructed_hdot}. Note that the matching conditions for perturbations are implicitly evaluated at $t = t^p_*$ (rather than $t^b_*$). In practice, it is advisable to rescale the variables in \eqref{eq:matching_re_im_dpsi} into dimensionless form (as in \eqref{eq:tilde_variables}) to improve numerical stability.

The matching conditions are simply ``initial'' conditions for the slow modes. Using these conditions, one can evolve the background and perturbation dynamics with EFT equations from \eqref{eq:psi_slow}-\eqref{eq:eta_slow}. As mentioned previously, these equations are non-oscillatory and therefore numerically stable, making them much easier to solve despite appearing more complicated than the exact equations. Once the slow modes of the wavefunction are obtained, the reconstruction of the exact fluid variables and metric perturbations follows as in Sec.~\ref{sec:reconstruction}.

\begin{figure}
    \centering
    \includegraphics[scale=0.7]{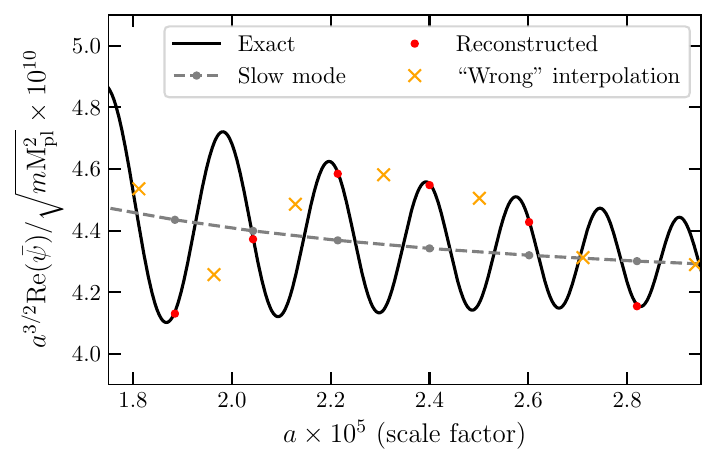}
    \caption{Real part of the wavefunction (rescaled), plotted for the axion field with $m = 10^{-24}~{\rm eV}$ and $\Omega_a = 0.2$. The black solid curve shows the function ${\rm Re}(\bar{\psi})$ computed to high precision, while other curves are derived from the EFT equations with the transition time at $a = 1.2 \times 10^{-5}$. Here, the gray dashed curve shows the slow mode ${\rm Re}(\bar{\psi}_s)$, constructed solely from the gray dots. The red dots indicate the reconstructed values of ${\rm Re}(\bar{\psi})$ at the same scale factors as the gray dots. The yellow crosses illustrate how this reconstructed function interpolates to new time points that do not coincide with the previous ones.}
    \label{fig:interpolation_issue}
\end{figure}

\subsection{Non-axion equations} \label{sec:non-axion_eqs}

Until now, we have only mentioned equations governing the wavefunction and metric perturbations. What about equations governing non-axion species? In principle, axion oscillations also induce oscillatory features on the fluid variables (density/pressure/velocity) of photon, baryons, CDM, etc. As these variables do not couple directly to the axion field (but they do couple indirectly via metric perturbations), one option is to reconstruct $\dot{h}$ and $\eta$ from $\dot{h}_s$ and $\eta_s$ (with \eqref{eq:reconstructed_hdot}, \eqref{eq:reconstructed_eta}) in the effective regime. The advantage of this approach is that there is no need to change anything else. However, the drawback is that the non-axion equations now contain fast oscillatory components throughout the whole cosmic history. When testing it in \texttt{CLAXions}, we observe that computations become significantly slower in the regions of parameter space where the axion field dominates (for large $\Omega_a$).

In order to remove the stiffness, we need an EFT version for {\it every} non-axion equation, which would require an enormous amount of work beyond the scope of the current study. Fortunately, up to the next-to-leading order in the corrections, one can prove that most EFT equations for non-axion perturbations retain the same form as their exact counterparts, with the exact variables simply replaced by the corresponding slow modes~\cite{Luu:2026las, Salehian:2020bon}. For example, the exact equation governing the density contrast of CDM, namely $\delta_c$, and its EFT version would look like
\begin{align}
    \dot{\delta}_c = \dfrac{\partial}{\partial t}\left( \dfrac{\delta\rho_c}{\bar{\rho}_c} \right) = -\dfrac{1}{2}\dot{h} \Hquad \rightarrow \Hquad \dfrac{\partial}{\partial t}\left( \dfrac{\delta\rho_{c,s}}{\bar{\rho}_{c,s}} \right) = -\dfrac{1}{2}\dot{h}_s. \label{eq:delta_cdm_eq}
\end{align}
This observation, together with the fact that the slow modes of non-axion fluid variables are also equal to the exact ones up to the next-leading order (as mentioned in \eqref{eq:nonaxion_variables}), implies that the non-axion equations of motion may be left unchanged. The only modification is to replace all metric-related variables with their slow-mode versions. Therefore, in \texttt{CLAxions}, we implement equation \eqref{eq:delta_cdm_eq} as
\begin{align}
    \dot{\delta}_c &= -\dfrac{1}{2}\dot{h} \quad {\rm when} \quad t \le t^p_*, \\
    \dot{\delta}_c &= -\dfrac{1}{2}\dot{h}_s \quad {\rm when} \quad t > t^p_*. \label{eq:delta_cdm}
\end{align}
Equations of other species should follow similarly\footnote{See App.~A of Ref.~\cite{Luu:2026las} for the EFT derivation of some non-axion equations.}, though they may involve different metric variables which are inferred from $\dot{h}_s$ and $\eta_s$.

\subsection{Interpolation} \label{sec:interpolation}

The second issue is related to numerical interpolation in an intermediate regime when $t^b_* < t < t^p_*$ for $k$ modes with $t^p_* \neq t_*^b$. In this time window, one still solve the exact perturbation equations, but they depend on the exact background quantities ({\it e.g.}, $\bar{\psi}$, $H$) that are no longer available. Therefore, the most natural solution is to reconstruct these background quantities  and use them in place of the exact ones. The question is {\it how} this reconstruction should be performed in conjunction with numerical interpolation.

For context, it is common practice in nearly all Boltzmann codes to solve the background dynamics first. The obtained results are then {\it cached} and {\it interpolated} whenever background quantities are needed later for perturbations or observables computation. In the $\Lambda$CDM model, this procedure works well because most background functions varying smoothly on the logarithmic timescale (typically measured by $\ln a$). Thus, we only need to compute and fill the interpolation table at a reasonable number of logarithmically-spaced discrete points in advance. However, we must exercise caution when applying the same interpolation procedure to oscillatory functions induced by axion dynamics. The reason is discussed below.

Figure \ref{fig:interpolation_issue} shows a concrete example with ${\rm Re}(\bar{\psi})$. Here, we see that the exact function of this variable (black curve) exhibits oscillations on a linear timescale of $m^{-1}$. Its corresponding slow mode (gray dashed curve), namely ${\rm Re}(\bar{\psi}_s)$, is a slowly-varying function on a logarithmic time scale. As such, the latter function can be represented just by {\it sparsely sampled} discrete points (gray dots), unlike the former\footnote{This is also why the EFT equations are non-stiff: the slow-mode variables can be integrated less frequently than the exact ones while still maintaining an accurate description of their evolution.}. Let us denote the scale factors at these points $\{a_i\}$. Now, imagine that we need to know ${\rm Re}(\bar{\psi})$ at new time points $\{a_j\}$ that are not overlapping with $\{a_i\}$. Given that the exact function (black curve) is not accessible in the effective regime and therefore cannot be used for interpolation, how should we proceed?

At first, it seems the most obvious way is to reconstruct ${\rm Re}(\bar{\psi})$ at $\{a_i\}$ and then interpolate this new function at $\{a_j\}$. The trouble is that, due to the sparse distribution of $\{a_i\}$, each point of the reconstructed ${\rm Re}(\bar{\psi})$ (red dots) could end up on a different oscillation cycle. As a result, ${\rm Re}(\bar{\psi})$ interpolated in this way (yellow crosses) completely fails to reproduce the exact function (black curve). One may argue that increasing the sample points $\{a_i\}$ may mitigate this problem. However, it would require a huge number of time points to resolve every oscillation cycle up to the present time, thereby defeating the purpose of the EFT formalism.

As it turns out, a better strategy is to interpolate the slow modes variables {\it before} reconstructing them. In the example above, ${\rm Re}(\psi_s)$ is smooth throughout the cosmic history, hence its interpolated values should accurately reflect the true function. The subsequent reconstruction then warrants a faithful reproduction of ${\rm Re}(\psi)$.

\section{EFT versus other methods} \label{sec:results}

\begin{figure*}[t]
    \centering
    \includegraphics[scale=0.8]{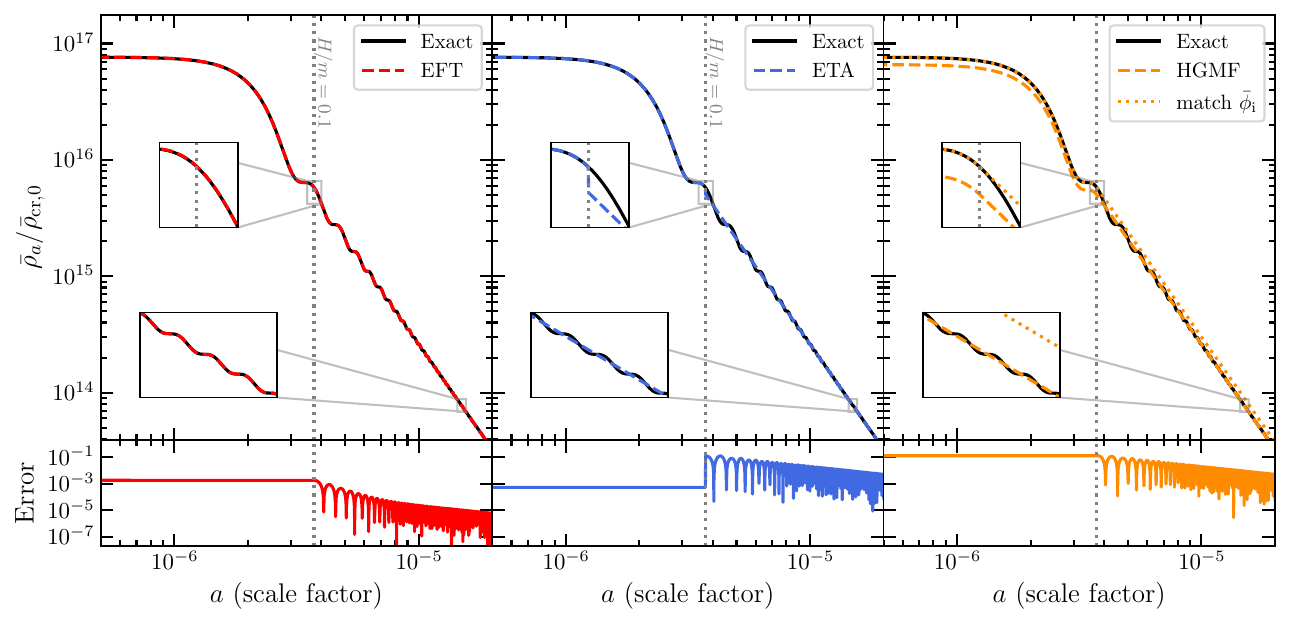}
    \caption{(Upper panels) Axion background density computed using the EFT (dashed red), ETA (dashed blue), and HGMF (dashed orange) approaches, compared with the exact result (solid black). The axion input parameters are chosen as $m = 10^{-23}~{\rm eV}$ and $\Omega_a = 0.26$ in all cases. We also set the transition time where $\epsilon_H = H/m = 0.1$ as shown by the (dotted gray) vertical line in each panel. The exact curve is obtained from a separate run with very late transition where $\epsilon_H = 0.0001$. For comparison, the right panel shows an additional curve (dotted orange) with $\bar{\phi}_i \simeq 2.4 \times 10^{17}~{\rm GeV}$ (instead of $\Omega_a = 0.26$), matching the initial field value of the black curve. The insets provide an enlarged view of the same indicated regions in each panel. (Lower panels) Fractional error of the background density in each approach with respect to its exact function. The error is defined by ${\rm Error} \equiv |1 - \bar{\rho}_a /\bar{\rho}^{\rm exact}_a|$ where $\bar{\rho}_a$ is the corresponding colored curves from the above panels and $\bar{\rho}^{\rm exact}_a$ indicates the black curves. In lower-right panel, $\bar{\rho}_a$ is taken from the above orange dashed curve (not the dotted one).}
    \label{fig:background_density}
\end{figure*}

\begin{figure*}[t]
    \centering
    \includegraphics[scale=0.8]{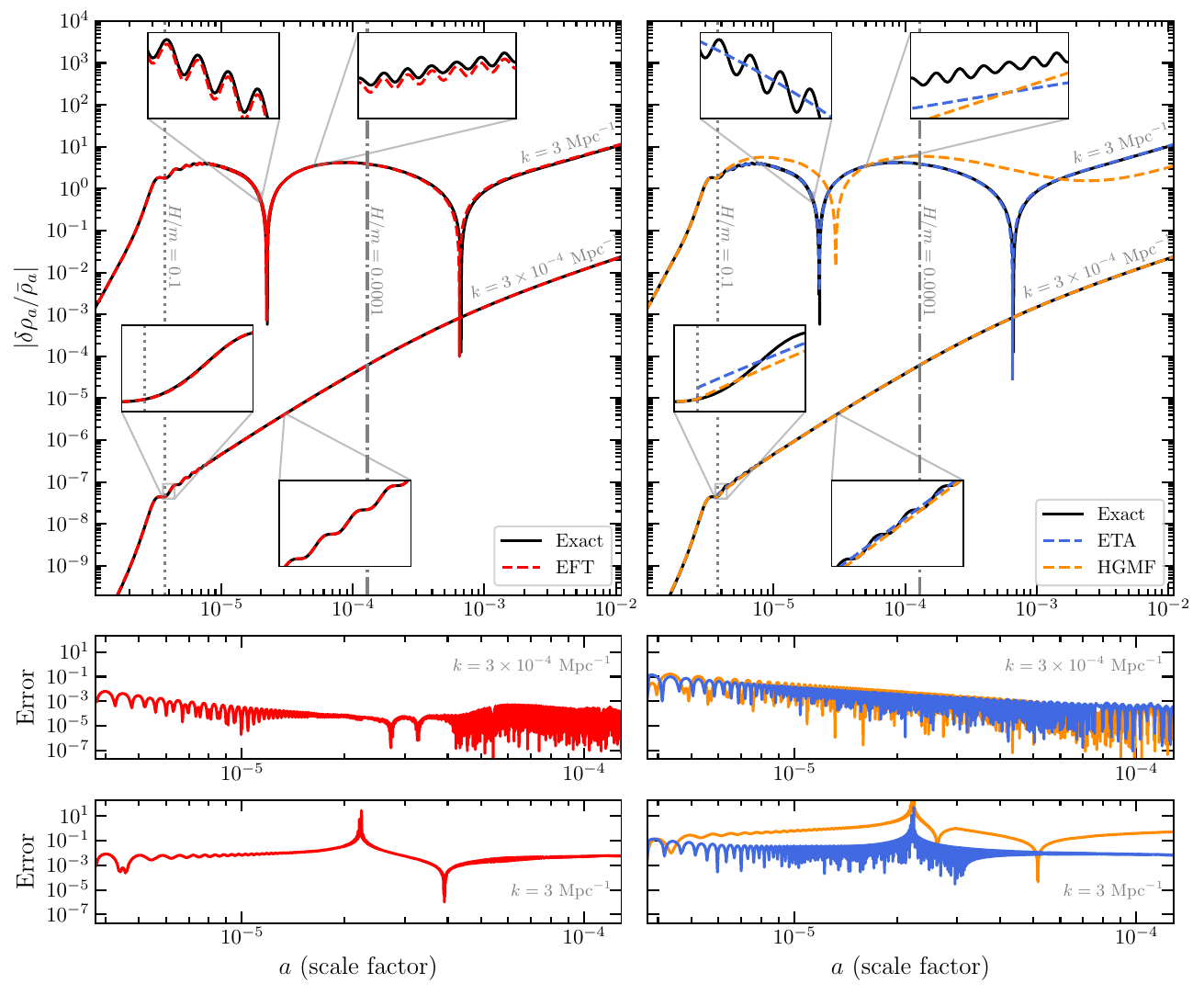}
    \caption{(Upper panels) Amplitude of the axion density contrast, computed using the EFT (dashed red), ETA (dashed blue), and HGMF (dashed orange) approaches, compared with the exact result (solid black). The input parameters and conventions are similar to those in Fig.~\ref{fig:background_density}. Although we show the same colored curves for two different wavenumbers $k = 3 \times 10^{-4}~{\rm Mpc}^{-1}$ (super-horizon) and $k = 3~{\rm Mpc}^{-1}$ (sub-horizon), they should be apparently separate in each panel. For the latter one, since $\epsilon_k \sim 0.2$ when $\epsilon_H = 0.1$, we intentionally choose $E_k > 0.25$ so that the transition times coincide among the three approaches (see discussions in Sec.~\ref{sec:transition_time}). Since the $a$-axes is more extended compared to Fig.~\ref{fig:background_density}, extra vertical lines (dashed-dot gray) are added to mark the transition time of the ``exact'' curves. Beyond these lines, they no longer display the exact result. (Lower panels) Fractional error of the density contrast in each approach with respect to its exact function. Each row shows the error of the indicated wavenumber, defined by ${\rm Error} \equiv |1 - \delta_a(k) /\delta^{\rm exact}_a(k)|$. The $a$-axes here only display the time range bounded by $H/m = 0.1$ and $H/m = 0.0001$ lines in the top panels. Note that the divergence peaks around $a \sim 2\times 10^{-5}$ for the high-$k$ mode in the bottom panels are not reliable because $\delta_a$ vanishes there.}
    \label{fig:perturbations_density}
\end{figure*}

In addition to the EFT method, \texttt{CLAxions} includes an alternative option to use the effective fluid approximation (EFA) used by \texttt{axionCAMB}~\cite{Hlozek:2014lca}, as well as \texttt{AxiECAMB}~\cite{Liu:2024yne}. These two cosmological codes were originally developed based on \texttt{CAMB}~\cite{Howlett:2012mh}. By translating their implementations into \texttt{CLASS}~\cite{Blas:2011rf}, we are able to perform a fair comparison with the new EFT approach presented in this study. For convenience, let us simply refer to the these other methods as ``HGMF'' (short for the authors' names) and the second one ``ETA'' (Effective Time Average)\footnote{``Effective time average'' refers to the time average of dynamical variables, a concept equivalent to slow modes in EFT.} from now on. The numerical implementation of these alternative methods is detailed in App.~\ref{app:EFA_implementation}.

As a reference, we set the fiducial axion model for comparison among different approaches as: $m = 10^{-23}~{\rm eV}$ and $\Omega_a = 0.26$. With this choice, the axion field replaces CDM as the dominant DM component, and $\Omega_c$ is set to be negligible. Other cosmological parameters are fixed to the best-fit values of the $\Lambda$CDM model constrained by Planck~\cite{Planck:2018vyg}: $\Omega_b = 0.02238, H_0 = 67.8~{\rm km}/s/{\rm Mpc}, A_s = 2.1 \times 10^{-9}, n_s = 0.966, z_{\rm reio} = 7.67$, and no exotic physics enabled. The transition time is identically chosen at $\epsilon_H = 0.1$, equivalently $a_* \sim 3.7 \times 10^{-6}$. After this moment, the slow-mode wavefunction is evolved instead of the exact wavefunction in the EFT approach (see Sec.~\ref{sec:numerical_implementation}) and when the effective fluid variables are evolved instead of the scalar field in the alternative approaches (see App.~\ref{app:EFA_implementation}). For reference, we set up the ``exact'' case, which is obtained by applying the EFT formalism with the transition time chosen sufficiently late at $\epsilon_H = 0.0001$, in order to achieve high-precision results.

\subsection{Background density} \label{sec:background_result}

Figure \ref{fig:background_density} compares the evolution of the axion background density in the EFT, ETA and HGMF approaches. From the upper left panel, it is clear that the density in the EFT approach (red dashed curve) matches the exact function (black solid curve) extremely well. The level of agreement becomes even more remarkable considering that the oscillatory features of the red curve after the transition time (gray dashed line) are reconstructed almost perfectly, as illustrated by the insets. In fact, the fractional error shown in the lower panel indicates that the EFT density deviates from the exact result by only a subpercent amount ($\lesssim  0.1\%$) at all times.

The ETA and HGMF approaches do not reconstruct the true density. Instead, they try to determine the density averaged over oscillation cycles. It is straightforward to see that, in both cases, the corresponding density (blue and orange dashed curves) redshift as dark matter after the transition time, {\it i.e.}, $\bar{\rho}_a \propto a^{-3}$. Nonetheless, these curves still capture the mean behavior of the exact function with relatively good precision. The fractional errors in these approaches after the transition time simply reflect the deviations of the true density from its average value, whose amplitude decays over time roughly as $a^{-3/2}$ from the WKB approximation~\cite{Ratra:1990me}.

A noticeable distinction between HGMF and ETA is that $\bar{\rho}_a$ remains continuous at the transition time in the former approach, whereas it becomes discontinuous in the latter. This discontinuity arises as a generic consequence of the ETA approach. It ensures that $\bar{\rho}_a$ always evolves to the correct present-day density (blue dashed curve) regardless of the chosen transition time. In contrast, in the HGMF approach, the density is likely to pick up a random phase within its current oscillation period at the transition time. As a result, one may either obtain an incorrect initial density $\bar{\rho}_{a,i}$ (and hence an incorrect $\bar{\phi}_i$) as we try to match $\Omega_a$ (orange dashed curve), or an incorrect $\Omega_a$ as we try to match $\phi_i$ (orange dotted curve) with the exact evolution.

\subsection{Density perturbations} \label{sec:perturbations_result}

Figure \ref{fig:perturbations_density} compares the axion density contrast, defined by $\delta_a \equiv \delta\rho_a/\bar{\rho}_a$, in the three approaches. We show the evolution of two modes: $k = 3 \times 10^{-4}~{\rm Mpc}^{-1}$ and $k = 3~{\rm Mpc}^{-1}$, which are chosen to represent super-horizon and sub-horizon fluctuations of the axion field.

Initially, axion perturbations follow the same adiabatic growth irrespective of the wavenumbers, as described by Eq.~\eqref{eq:ics_perturbations}. Once they become dynamical (around the transition time), their growth behavior changes depending on the corresponding wavenumber. Large-scale perturbations behave similarly to those of CDM, whereas small-scale perturbations are suppressed by Jeans instability~\cite{Marsh:2015xka}. These features are clearly visible in the evolution of $\delta_a$ for all three approaches in Fig.~\ref{fig:perturbations_density}.

As with the background density, the true axion density contrast is closely reproduced by the EFT approach (red dashed curves) for both wavenumbers. From the insets in the upper-left panel of Fig.~\ref{fig:perturbations_density}, we observe that the reconstructed axion oscillations, despite their small amplitude, match the exact function down to individual oscillation periods. Even so, the EFT results now exhibit slight deviations from the exact solution, particularly for the high-$k$ mode. There are several reasons behind this small disagreement. First, the EFT equations at the perturbative level \eqref{eq:dpsi_slow} were derived up to lower-order corrections compared to those at the background level \eqref{eq:psi_slow}. Second, the error in the computation of $\delta_a$ also inherits not only the error associated with $\delta\rho_a$, but also that of $\bar{\rho}_a$. Third, the dynamics of small-scale perturbations are more complex, as the axion oscillations become convoluted with Jeans oscillations induced by the metric perturbations. 

That said, these results are sufficiently accurate for the first demonstration of this approach as the reconstructed $\delta_a$ still matches the exact function at the percent level, which is indicated by the fractional error shown in the lower panel. It shows that the true axion dynamics can be already well captured by the EFT formalism at the lowest-order corrections. We expect a dramatic improvement once higher-order corrections for $\delta\psi_s$ are included in both the equations of motion \eqref{eq:dpsi_slow} and the matching conditions \eqref{eq:reconstructed_dpsi}, following the procedure in~\cite{Luu:2026las}. However, doing so may also require introducing corrections of the same order in the non-axion equations \eqref{eq:delta_cdm}, as well as in the approximation of their slow modes \eqref{eq:nonaxion_variables}, in order to maintain consistency. Given that this is by no means a trivial task, we defer the problem to future work.

\begin{figure}
    \centering
    \includegraphics[scale=0.7]{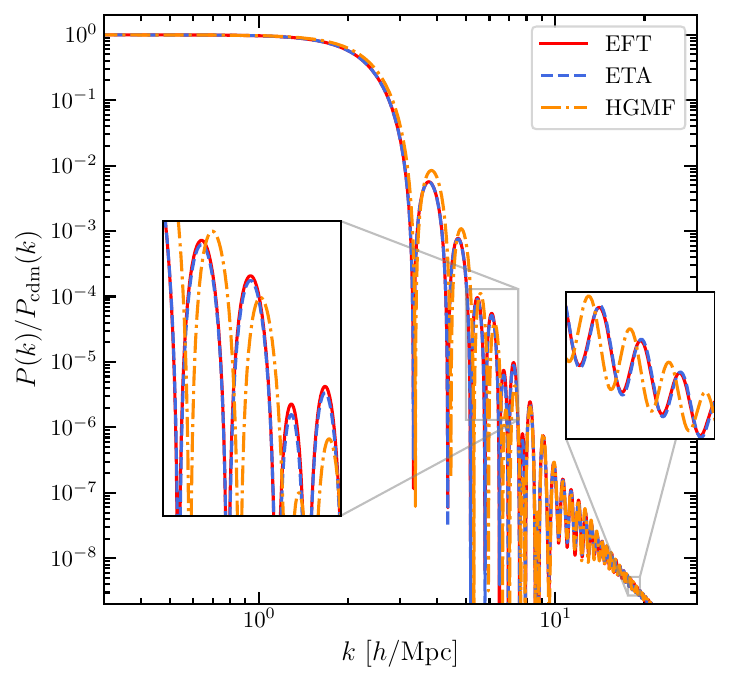}
    \caption{Axion matter power spectrum computed using the EFT (dashed red), ETA (solid blue), and HGMF (dashed-dotted orange) approaches. Input parameters are chosen as $m = 10^{-23}~{\rm eV}$ and $\Omega_a = 0.26$, whereas the precision parameters are $E_H = E_k = 0.1$. All spectra are shown relative to the ``pure'' CDM case, denoted as $P_{\rm cdm}(k)$, where $\Omega_a =0$ and $\Omega_{\rm c} = 0.26$. Note that there is no ``exact'' curve as in Fig.~\ref{fig:background_density} and Fig.~\ref{fig:perturbations_density} because the power spectrum is not given with respect to time, and therefore cannot be integrated exactly for this axion mass.}
    \label{fig:power_spectrum}
\end{figure}

By comparison, from the upper-right panel, $\delta_a$ computed using the ETA and HGMF approaches (blue and orange dashed curves) provides an overall good fit for the low-$k$ mode, but faces similar challenges for the high-$k$ mode. In the case of HGMF, the abrupt transition to the effective fluid description even results in an incorrect evolution of the high-$k$ mode after the transition time. Such a mismatch in the density perturbations may propagate into the computation of cosmological observables, such as the matter power spectrum as illustrated in Fig.~\ref{fig:power_spectrum}.

Here, one observes noticeable deviations of $P(k)$ on small scales in the HGMF approach (orange dashed-dotted curve) compared to the other two (red solid and blue dashed curves), likely due to the above-mentioned disagreement in high-$k$ modes of $\delta_a$. On the other hand, the power spectra computed from ETA and EFT are in excellent agreement, suggesting that the reconstructed oscillations of $\delta_a$ in the latter approach do not significantly affect this observable, at least for the field within the Fuzzy Dark Matter mass range ($m = 10^{-23}~{\rm eV}$ in this case). It would be interesting to extend this analysis to axions with various masses and density fractions to determine whether such oscillatory features could induce significant deviations in the matter power spectrum that are resolvable by future Lyman-$\alpha$ and $21$-cm experiments, thereby placing tighter constraints on axion physics.

\subsection{Slow mode and time average} \label{sec:slow-mode_result}

\begin{figure}
    \centering
    \includegraphics[scale=0.71]{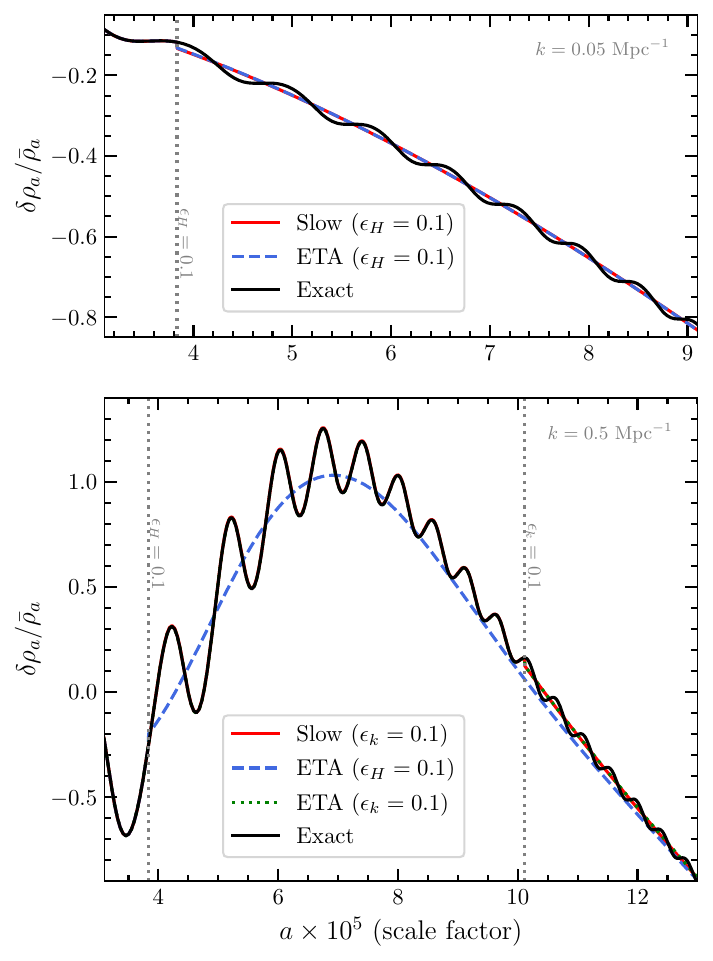}
    \caption{Axion density contrast around the transition time, computed for $k = 0.05~{\rm Mpc}^{-1}$ (upper) and $k = 0.5~{\rm Mpc}^{-1}$ (lower). The input parameters are $m = 10^{-25}~{\rm eV}$ and $\Omega_a = 0.03$. The text inside parenthesis after each label, {\it i.e.}, ``$(\epsilon_H = 0.1)$'' and ``$(\epsilon_k = 0.1)$'', denotes when the transition time occurring. These moments are also marked by the vertical gray lines in each panel. The red solid curves display the exact function $\delta\rho_a/\bar{\rho}_a$ before the transition time, and its associated slow mode $\delta\rho_{a,s}/\bar{\rho}_{a,s}$ after the transition time, with the EFT approach. The remaining curves illustrate the exact density contrast (black solid) and those computed by the ETA approach (blue dashed and green dotted), as in Fig.~\ref{fig:perturbations_density}.}
    \label{fig:slow_mode}
\end{figure}

Finally, let us briefly discuss the slow modes in EFT and the effective variables in ETA since they are supposed to represent the true time average of the exact quantities. Figure \ref{fig:slow_mode} demonstrates how they compare in case of the axion density contrast with the input parameters $m = 10^{-25}~{\rm eV}$ and $\Omega_a = 0.03$.

For a long wavelength mode such as $k = 0.05~{\rm Mpc}^{-1}$ (upper panel), the transition time in the EFT approach coincides with that of the ETA approach at $\epsilon_H = 0.1$, making the comparison straightforward. In this case, $\delta^{\rm ETA}_a$ (blue dashed) and $\delta^{\rm EFT}_a$ (red solid) are in perfect agreement after the transition time, and both accurately describe the mean of the exact $\delta_a$ function (black solid).

For a shorter wavelength mode such as $k = 0.5~{\rm Mpc}^{-1}$ (lower panel), the comparison becomes tricky, as $\epsilon_k$ has not yet reached a sufficiently small value at $\epsilon_H = 0.1$, and the EFT transition time must therefore be chosen later than the ETA one (see Sec.~\ref{sec:transition_time}). Nevertheless, one can already see that $\delta^{\rm ETA}_a$ with a transition at $\epsilon_H = 0.1$ (blue dashed) fails to reproduce the true average of $\delta_a$. Meanwhile, $\delta^{\rm EFT}_a$ with a later transition at $\epsilon_k = 0.1$ (red solid) accurately match it. If the transition time for $\delta^{\rm ETA}_a$ is also chosen at $\epsilon_k = 0.1$\footnote{We need to choose the transition time at $\epsilon_H \sim 0.01566$ for the ETA approach, so that $\epsilon_k \sim 0.1$ there.} (green dotted), the two effective descriptions are once again in agreement.

This result leads to an interesting realization. From a technical standpoint, ETA is considerably more straightforward to implement than EFT, given that both methods enforce the effective variables to represent true time-averaged quantities. For instance, the criteria for selecting the transition time are less stringent in the former approach, as a single condition is sufficient for both the background and perturbation equations (see App.~\ref{app:EFA_implementation}). However, when dealing with challenging cases such as the high-$k$ mode discussed above, a more conservative transition time must still be adopted to ensure the accuracy of the ETA approach. In this regard, the more stringent transition time requirement of the EFT approach for perturbations effectively serves as a natural safeguard.

Interestingly, by removing the reconstruction step and treating the slow-mode variables directly as the exact ones, one can construct an approach that is equivalent to ETA but based on the EFT formalism. In fact, doing so even eliminates the need to handle non-axion equations, as mentioned in Sec.~\ref{sec:non-axion_eqs}. The advantage of implementing the effective fluid approximation within the EFT framework is that the resulting equations of motion can be readily generalized to more complex axion theories~\cite{Luu:2026las}, such as those involving self-interactions. The same generalization would be considerably more difficult in the ETA approach~\cite{Passaglia:2022bcr}, as one would need to recalibrate the equation of state and sound speed for effective fluid variables with new theories.

\section{Conclusion} \label{sec:conclusion}

In this work, we have used numerical methods to establish an EFT with relativistic corrections as a robust framework for modeling cosmological axions. Unlike previous approaches using effective fluid approximation with cycle-averaging, EFT is the only formalism capable of reconstructing the relativistic oscillations of dynamical variables. As a result, it naturally avoids discontinuities in the resulting fluid quantities while providing a highly accurate description of the effective axion fluid.

Whether these reconstructed oscillatory features are observationally detectable remains an open and intriguing question. One area of research that may see an immediate benefit is the search for ultralight axions using pulsar timing arrays~\cite{Khmelnitsky:2013lxt}. In this context, the observed signals in the pulsar timing residuals receive deterministic contribution from the background oscillations of the axion field. The amplitude and phase of these oscillations are typically assumed to be uncorrelated, owing to the lack of knowledge about their exact values at the present day. In principle, however, a single initial field value uniquely determines both the amplitude and phase of the axion oscillations at any subsequent time through the field evolution. Therefore, by reconstructing the axion field to high precision, it may be possible to establish the exact correlation between these two quantities.

On the technical side, there is the possibility of extending the EFT approach to scalar field theories with a general potential. Since a first attempt on the theoretical front has been initiated recently~\cite{Modirzadeh:2025gjd}, we are motivated to develop a systematic approach for the equivalent numerical implementation of that general case for future work. In the meantime, we hope the current EFT approach will find broader applications across astrophysics and cosmology.

\begin{acknowledgments}
We are thankful to Mohammad Hossein Namjoo and Patrick Fitzpatrick for helpful comments and conversations. We also thank Luna Zagorac, Nathan Musoke, Mark Neyrinck, David Kaiser, Risa Wechsler, and Priya Natarajan for encouraging conversations along the way. We thank the administrative and facilities staff at the University of New Hampshire including Katie Makem-Boucher and Michelle Mancini. HNL and CPW's contributions to this project were supported by DOE Grant DE-SC0025365, and we gratefully acknowledge the work of the DOE program officers who support the management of this grant.
\end{acknowledgments}

\appendix

\section{Effective fluid approximation} \label{app:EFA_implementation}

This appendix outlines the numerical implementation of the effective fluid approximation used in the HGMF~\cite{Hlozek:2014lca} and ETA~\cite{Liu:2024yne} approaches. Readers seeking a more thorough discussion are referred to the respective original papers. Here, our goal is to provide a brief summary of the most relevant equations as they are implemented in \texttt{CLAxions}, while also standardizing the notation used in those studies to be consistent with that of the EFT equations in the main text.

Regardless of whether one is dealing with HGMF or ETA, the equations governing the axion field in the exact regime are identical. Assuming the axion field is decomposed as $\phi(t,\bm{x}) = \bar{\phi}(t) + \delta\phi(t,\bm{x})$, we can derive
\begin{align}
    \ddot{\bar{\phi}} + 3H\dot{\bar{\phi}} + m^2\bar{\phi} = 0, \label{eq:phi_ax}
\end{align}
at the background level and
\begin{align}
    \ddot{\delta\phi} + 3H\delta\dot{\phi} + \left(\dfrac{k^2}{a^2} + m^2\right)\delta\phi + \dfrac{1}{2}\dot{\bar{\phi}}\dot{h} = 0 \label{eq:dphi_ax}
\end{align}
at the perturbation level (synchronous gauge)~\cite{Hu:2003hjx}. In fact, these two equations are also equivalent to Eqs.~\eqref{eq:psi_exact_background} and \eqref{eq:dpsi_exact_perturbations} for $\bar{\psi}$ and $\delta\psi$ in the EFT approach. Therefore, all three approaches are expected to yield the same results before the background transition time, as confirmed by the axion density shown in Fig.~\ref{fig:background_density} and Fig.~\ref{fig:perturbations_density}.

The axion fluid variables in this regime, including the density, pressure and velocity perturbations, can be expressed in terms of $\phi$ as
\begin{align}
    &\bar{\rho}_a = \dfrac{1}{2}\dot{\bar{\phi}}^2 + \dfrac{1}{2}m^2\phi^2, \quad \bar{p}_a = \dfrac{1}{2}\dot{\bar{\phi}}^2 - \dfrac{1}{2}m^2\phi^2, \\
    &\delta\rho_a = \dot{\bar{\phi}}\dot{\delta\phi} + m^2\bar{\phi}\delta\phi, \quad \delta p_a = \dot{\bar{\phi}}\dot{\delta\phi} - m^2\bar{\phi}\delta\phi, \nonumber \\
    &\delta U_a = - \dot{\bar{\phi}}\delta\phi.
\end{align}

The transition time in HGMF and ETA is determined by a single condition, $\epsilon_H = H/m < E_H$. By default, $E_H = 0.1$ in \texttt{CLAxions}, consistent with the recommended value in \texttt{AxiECAMB} for the ETA approach~\cite{Liu:2024yne}. This is a more conservative choice compared to the default value of $E_H = 1/3$ used in \texttt{axionCAMB}~\cite{Hlozek:2014lca} for the HGMF approach.

The numerical implementation follow exactly the first four steps mentioned in Sec.~\ref{sec:numerical_implementation}. We evolve Eqs.\eqref{eq:phi_ax} and \eqref{eq:dphi_ax} with initial conditions given by Eqs.~\eqref{eq:ics_background} and \eqref{eq:ics_perturbations} until the transition time, at which point the matching conditions are applied and the evolution continues with the effective equations. Only these last two steps differ between the two approaches, so we discuss them separately below.

\subsection{EFA with HGMF}

Technically, there is no matching condition in the HGMF approach. Instead, one continues to evolve the axion fluid variables from their last values (at the transition time) in the exact regime into the effective regime. Note that, unlike the EFT approach, HGMF and ETA treat the effective fluid variables {\it as if} they were the exact ones. The matching conditions simply read
\begin{align}
    &\bar{\rho}_{a,*} = \bar{\rho}_a(t_*), \quad \delta_{a,*} = \dfrac{\delta\rho_a(t_*)}{\bar{\rho}_a(t_*)}, \\
    &u_{a, *} =  - \dfrac{k}{a_*}\dfrac{\delta U_a(t_*) }{\bar{\rho}_a(t_*)} \Hquad \text{where} \Hquad a_* = a(t_*).
\end{align}
$u_a$ as defined above is called the \textit{axion heat flux}~\cite{Hlozek:2014lca}.

After the transition time, the effective fluid are characterized by the effective fluid variables rather than the field $\phi$. At the background level, the effective axion fluid is treated as collisionless dark matter. The effective background density and pressure are therefore given by
\begin{align}
    \bar{\rho}_a = (\bar{\rho}_{a,*}a^3_*)a^{-3}, \quad \bar{p}_a = 0.
\end{align}
Meanwhile, the perturbative axion fluid is governed by the following equations
\begin{align}
    &\dot{\delta}_a = -\dfrac{k}{a}u_a - 3Hc^2_s\delta_a - \dfrac{9a}{k}H^2c^2_s u_a - \dfrac{1}{2}\dot{h}, \label{eq:delta_a} \\
    &\dot{u}_a = - Hu_a + \dfrac{k}{a}c_s^2\delta_a + 3Hc_s^2u_a. \label{eq:heat_a}
\end{align}
Here, $c_s$ denotes the axion sound speed computed in the {\it axion-comoving} gauge\footnote{This is the coordinate system comoving with the ``averaged'' axion fluid (again, not the exact one) where $\braket{v_a} = 0$.}, given by
\begin{align}
    c_s^2 = \dfrac{k^2}{4m^2a^2}\left( 1 + \dfrac{k^2}{4m^2a^2} \right)^{-1},
\end{align}
where the derivation has been detailed in \cite{Hwang:2009js}. The evolution of the effective fluid is then determined by solving Eqs.~\eqref{eq:delta_a} and \eqref{eq:heat_a}, with $\delta_{a,*}$ and $u_{a,*}$ serving as initial conditions at the transition time. 

Lastly, the effective pressure perturbations are computed by invoking the definition $c^2_s \equiv \braket{\delta p_a}/\braket{\delta\rho_a}$ in the axion-comoving gauge, followed by a gauge transformation back to the synchronous gauge (see App.~A in \cite{Poulin:2018dzj})
\begin{align}
    \delta p_a = c_s^2\delta\rho_a + \dfrac{3aH}{k}c_s^2\bar{\rho}_au_a.
\end{align}
Interestingly, we notice that the effective $\bar{p}_a$ and $\delta p_a$ are not required to match the exact solutions at the transition time. As a result, they are likely to be discontinuous there, potentially sourcing spurious Jeans oscillations.

\subsection{EFA with ETA}

Conceptually, the implementation of ETA is similar to that of HGMF. The main distinction lies in the matching conditions where the effective fluid quantities are strongly dependent on the exact solution at the transition time. The main idea is to introduce two auxiliary field variables $\varphi_c, \varphi_s$ satisfying~\cite{Passaglia:2022bcr}
\begin{align}
    \bar{\phi}(\tilde{t}) = \varphi_c(\tilde{t})\cos(\tilde{t}-\tilde{t}_*) + \varphi_s(\tilde{t})\sin(\tilde{t}-\tilde{t}_*),
\end{align}
where $\tilde{t} = mt$. Since the original (real) field $\phi$ is now described by two degrees of freedom, there is a freedom in choosing the matching conditions such that $\varphi_c$ and $\varphi_s$ vary only on the Hubble timescale. To clarify, the exact equations governing $\varphi_c$ and $\varphi_s$ are still oscillatory and cannot be solved all the way to $z = 0$. However, certain conditions strongly suppress the amplitude of oscillations in these two variables. For instance, we may set~\cite{Passaglia:2022bcr}
\begin{align}
    \dfrac{\varphi''_{c,*}}{\varphi'_{c,*}} = \dfrac{\varphi''_{s,*}}{\varphi'_{s,*}} = -\dfrac{3\braket{H}_*}{2m} + \dfrac{\braket{H}'_*}{\braket{H}_*}, \label{eq:eta_cond_background}
\end{align}
where $X' = \partial X/\partial \tilde{t}$ and $\braket{H}'_*$ is $\braket{H}'$ computed at $t_*$. Note that $\braket{H}$ denotes the time average of the Hubble function, which in general differs from $H$. One may think of $\braket{X}$ as the equivalence of $X_s$ in the EFT approach.

Without going into further detail (of which we refer to \cite{Liu:2024yne}), the conditions \eqref{eq:eta_cond_background}, together with the matching of $\bar{\phi}$ and $\bar{\phi}'$ with the auxiliary variables at $t_*$, yield
\begin{align}
    \varphi_{c,*} &= \bar{\phi}_*, \quad \varphi_{s,*} = \bar{\phi}'_* - \varphi'_{c,*}, \\
    \varphi'_{c,*} &= - \dfrac{3\tilde{H}_*\left[ 2\bar{\phi}_* + (\mathcal{A}_* + 3\tilde{H}_*)\bar{\phi}'_* \right]}{\mathcal{A}_*^2 + 3\tilde{H}_*\mathcal{A}_* + 4}, \\
    \varphi'_{s,*} &= \dfrac{3\tilde{H}_*(\mathcal{A_*}\bar{\phi}_* - 2\bar{\phi}'_*)}{\mathcal{A}_*^2 + 3\tilde{H}_*\mathcal{A}_* + 4},
\end{align}
where we have introduced
\begin{align}
    \tilde{H}_* = \dfrac{H_*}{m}, \quad \mathcal{A}_* = -\dfrac{3\braket{H}_*}{2m} + \dfrac{\braket{H}'_*}{\braket{H}_*}. \label{eq:A_eta}
\end{align}
The time-averaged axion density and pressure are then obtained via cycle-averaging as
\begin{align}
    &\braket{\bar{\rho}_a}_* = \dfrac{m^2}{4}\left[ \varphi'^2_{c,*} + \varphi'^2_{s,*} \right. \nonumber \\ &\hspace{1.2cm} \left. +  2\left( \varphi_{c,*}^2 + \varphi_{s,*}^2 - \varphi_{c,*} \varphi'_{s,*} + \varphi_{s,*}\varphi'_{c,*} \right)  \right], \label{eq:rho_eta} \\
    &\braket{\bar{p}_a}_* = \braket{\bar{\rho}_a}_* - \dfrac{m^2}{2}(\varphi_{c,*}^2 + \varphi_{s,*}^2). \label{eq:p_eta}
\end{align}
While these final expressions may appear complicated, they are completely deterministic once the exact solution at $t_*$ is specified. The only remaining unknown is $\mathcal{A}_*$ with $\braket{H}_*$ and $\braket{H}'_*$ defined by replacing $\bar{\rho}_a$ with $\braket{\bar{\rho}_a}$ as follows
\begin{align}
    3\Mpl^2\braket{H}^2_* &= \braket{\bar{\rho}_a}_* + \bar{\rho}_{\as,*}, \label{eq:H_eta} \\
    2\Mpl^2\braket{H}'_* &= - \braket{\bar{\rho}_a}_* - \braket{\bar{p}_a}_* - \bar{\rho}_{\as,*} - \bar{p}_{\as,*}. \label{eq:Hprime_eta}
\end{align}
To resolve the circular dependence of these quantities on $\braket{\bar{\rho}_a}_*$ and $\braket{\bar{p}_a}_*$, one can employ the same iterative procedure as in the EFT approach. We begin by setting $\braket{H}_* = H_*$ and $\braket{H}'_* = H'_*$ as initial guesses, then compute $\mathcal{A}_*,\braket{\bar{\rho}_a}_*, \braket{\bar{p}_a}_*$ using \eqref{eq:A_eta}, \eqref{eq:rho_eta}, \eqref{eq:p_eta}, then subsequently substitute the results back into \eqref{eq:H_eta}, \eqref{eq:Hprime_eta} to obtain updated values of $\braket{H}_*$ and $\braket{H}'_*$. These steps are repeated until $\braket{\bar{\rho}_a}_*$ and $\braket{\bar{p}_a}_*$ converge, usually after a few iterations.

Recall that in the ETA approach, the effective fluid variables are treated as the exact ones. That means, the distinction between the exact solution $X$ and its time average $\braket{X}$ is only relevant at the transition time. Once the system enters the effective regime, the two become indistinguishable. Thus, $\braket{\bar{\rho}_a}_*$ as given in \eqref{eq:rho_eta} also serves as the initial condition for the equation
\begin{align}
    \dot{\bar{\rho}}_a = - 3H(1 + w_a)\bar{\rho}_a,
\end{align}
governing the effective evolution of $\bar{\rho}_a = \braket{\bar{\rho}_a}$ for $t > t_*$. Here, the effective axion equation of state $w_a$ is a time-dependent function calibrated to accurately capture the relation between the mean density and pressure. It is parametrized as
\begin{align}
    w_a = \dfrac{\braket{\bar{p}_a}_*}{\braket{\bar{\rho}_a}_*\tilde{H}_*^2}\tilde{H}^2, \label{eq:eos_eta}
\end{align}
where the numerical coefficient in front of $\tilde{H}^2$ is chosen such that $w_a(t_*) = \braket{\bar{p}_a}_*/\braket{\bar{\rho}_a}_* \sim 3/2$. The effective pressure can  therefore be computed as $\bar{p}_a = w_a\bar{\rho}_a$, such that $\bar{p}_a(t_*) = \braket{\bar{p}_a}_*$

The ETA implementation at the perturbation level follows the same general structure. First, the matching conditions read
\begin{align}
    \delta\varphi_{c,*} &= \delta\phi_*, \quad \delta\varphi_{s,*} = \delta\phi'_{c,*} - \delta\varphi'_{c,*}, \\
    \delta\varphi'_{c,*} &= - \dfrac{2\mathcal{B}_* + (\mathcal{A}_* + 3\tilde{H}_*)\mathcal{C}_*}{2\left(\mathcal{A}^2_* + 3\tilde{H}_*\mathcal{A}_* + 2\tilde{k}_* + 4 \right)}, \\
    \delta\varphi'_{s,*} &= \dfrac{\mathcal{A}_*\mathcal{B}_* - (2 + \tilde{k}^2_*)\mathcal{C}_*}{2\left(\mathcal{A}^2_* + 3\tilde{H}_*\mathcal{A}_* + 2\tilde{k}_* + 4 \right)},
\end{align}
where we have introduced the following notations
\begin{align}
    \tilde{k}_* &= \dfrac{k^2}{m^2a_*^2}, \\
    \mathcal{B}_* &= (\varphi_{c,*} - \varphi'_{s,*})h'_* - 2\tilde{k}_*\delta\phi'_* + 6\tilde{H}_*\delta\phi_*, \\
    \mathcal{C}_* &= (\varphi_{s,*} + \varphi'_{c,*})h'_* + 2\tilde{k}_*\delta\phi_* + 6\tilde{H}_*\delta\phi'_*,
\end{align}
besides $\delta\varphi_c$ and $\delta\varphi_s$, which are just the perturbations of $\varphi_c$ and $\varphi_s$. The time-averaged fluid variables are now
\begin{align}
    &\braket{\delta\rho_a}_* = \dfrac{m^2}{2}\left[ (\varphi'_{s,*} - \varphi_{c,*})\delta\varphi'_{s,*} + (\varphi'_{c,*} + \varphi_{s,*})\delta\varphi'_{c,*} \right. \nonumber \\ 
    &\hspace{0.5cm} \left. + (\varphi'_{c,*} + 2\varphi_{s,*})\delta\varphi_{s,*} - (\varphi'_{s,*} - 2\varphi_{c,*})\delta\varphi_{c,*} \right], \label{eq:delta_rho_eta} \\
    &\braket{\delta p_a}_* = \braket{\delta\rho_a}_*  - m^2(\varphi_{c,*}\delta\varphi_{c,*} + \varphi_{s,*}\delta\varphi_{s,*}), \label{eq:delta_p_eta} \\
    &\braket{\delta U_a}_* = - \dfrac{m}{2}\left[ (\varphi'_{c,*} + \varphi_{s,*})\delta\varphi_{c,*} + (\varphi'_{s,*} - \varphi_{c,*})\delta\varphi_{s,*} \right]. \label{eq:delta_U_eta}
\end{align}
In addition, the time average of the metric perturbations must be computed to serve as their matching conditions
\begin{align}
    \braket{\eta}_* = \eta_* - \left[\dfrac{a^2_*}{2k^2}\braket{H}_*(\dot{h}_* + 6\dot{\eta}_*) - \eta_* - \mathcal{E}_*\right]\dfrac{\mathcal{D}_*}{1 - \mathcal{D}_*}, \label{eq:eta_eta}
\end{align}
where
\begin{align}
    \mathcal{D}_* &= - \dfrac{k}{3a_*\braket{H}_*}, \\
    \mathcal{E}_* &= \dfrac{a_*^2}{2\Mpl^2k^2}\left(\braket{\delta\rho_a}_* + \delta\rho_{\as,*} \right. \nonumber \\ &\hspace{1.8cm} \left. - 3\braket{H}_*\braket{\delta U_a}_* -3\braket{H}_* \delta U_{\as,s} \right).
\end{align}
If needed, the time-averaged variable $\braket{\dot{h}}_*$ can be solved from the ``00'' Einstein equation
\begin{align}
    \Mpl^2\left(\braket{H}_*\braket{\dot{h}}_* - \dfrac{2k^2}{a^2_*}\braket{\eta}_*\right) = \braket{\delta\rho_a}_* + \delta\rho_{\as,*}. \label{eq:hdot_eta}
\end{align}
Fortunately, no iteration is required here, as most of the time-averaged quantities have already been computed at the background level.

As in the case of HGMF, the effective equations governing the perturbations can be derived as
\begin{align}
    &\dot{\delta}_a = -\dfrac{k}{a}u_a - 3H(c^2_s - w_a)\delta_a \nonumber \\ &\hspace{1cm} - \dfrac{9a}{k}H^2(c^2_s - c^2_{\rm ad}) u_a - \dfrac{(1+w_a)}{2}\dot{h}, \label{eq:delta_eta} \\
    &\dot{u}_a = - H(1 - 3w_a)u_a + \dfrac{k}{a}c_s^2\delta_a + 3H(c_s^2 - c^2_{\rm ad})u_a, \label{eq:heat_eta}
\end{align}
where the effective adiabatic sound speed $c^2_{\rm ad}$ and the effective adiabatic sound speed $c^2_s$ are given by
\begin{align}
    c^2_s &= \dfrac{ \left(\sqrt{1 + \tilde{k}} - 1 \right)^2}{\tilde{k}} + \dfrac{5}4\tilde{H}^2, \label{eq:cs_eta} \\
    c^2_{\rm ad} &= w_a - \dfrac{w'_a}{3\tilde{H}(1+w_a)} \Hquad \text{where} \Hquad w'_a = \dfrac{2\braket{\bar{p}_a}_*}{\braket{\bar{\rho}_a}_*\tilde{H}_*^2}\tilde{H}\tilde{H}'. \label{eq:cad_eta}
\end{align}
The axion effective pressure perturbation can then be evaluated with
\begin{align}
    \delta p_a = c_s^2\delta\rho_a + \dfrac{3aH}{k}(c_s^2-c^2_{\rm ad})\bar{\rho}_au_a. \label{eq:delta_p_eta}
\end{align}
Note that we set the initial conditions for Eqs.~\eqref{eq:delta_eta} and \eqref{eq:heat_eta} such that
\begin{align}
    \delta_{a,*} = \dfrac{\braket{\delta\rho_a}_*}{\braket{\bar{\rho}_a}_*}, \quad u_{a,*} = - \dfrac{k}{a_*}\dfrac{\braket{\delta U_a}_*}{\braket{\bar{\rho}_a}_*},
\end{align}
with $\braket{\delta\rho_a}_*$ and $\braket{\delta U_a}_*$ substituted from \eqref{eq:delta_rho_eta} and \eqref{eq:delta_U_eta}, respectively. The matching conditions for the Einstein equations are similarly set, with the effective $\eta$ and $h$ initialized to $\braket{\eta}_*$ and $\braket{\dot{h}}_*$ as given in \eqref{eq:eta_eta} and \eqref{eq:hdot_eta}.

Compared to HGMF, the ETA implementation follows the same effective equations. The key difference is that the effective fluid now has a definitive equation of state and adiabatic sound speed, whereas they are assumed to vanish in HGMF. Interestingly, $w_a$ and $c_{\rm ad}$ as given by \eqref{eq:eos_eta} and \eqref{eq:cad_eta} depend on variables evaluated at $t_*$, implying that the effective equations will be modified each time a different value of the transition time is chosen.

\section{Power-series initial conditions} \label{app:power_series_ICs}

The axion initial conditions expressed as power series in conformal time have been discussed in the literature, {\it e.g.}, in the Appendices of \cite{Hlozek:2014lca} and \cite{Smith:2019ihp}. Realizing the procedures described in those works are not straightforward to follow, we provide a more explicit and self-contained derivation of these initial conditions here.

Let us begin at the background level. Initial conditions are typically set at very early times, when the universe is still radiation-dominated. In this regime, the Hubble function and scale factor are simply given by
\begin{align}
    H = \mathcal{C} a^{-2} \Hquad \rightarrow \Hquad a = \mathcal{C}\tau,
\end{align}
where $\mathcal{C} = H_0\sqrt{\Omega_r}$. Introducing the dimensionless conformal time $\tilde{\tau} = \mathcal{C}\tau$, we can write
\begin{align}
    a = \tilde{\tau} \quad \text{and} \quad H = \mathcal{C}\tilde{\tau}^{-2}. \label{eq:aH_ic}
\end{align}
Substituting $a$ and $H$ from \eqref{eq:aH_ic} into the axion background equation \eqref{eq:phi_ax} and converting the time derivatives to those with respect to $\tilde{\tau}$ yields
\begin{align}
    \tilde{\tau}\dfrac{\partial^2\bar{\phi}}{\partial\tilde{\tau}^2} + 2\dfrac{\partial\bar{\phi}}{\partial\tilde{\tau}} + \dfrac{m^2}{\mathcal{C}^2}\tilde{\tau}^3\bar{\phi} = 0. \label{eq:phi_ax_tau}
\end{align}

We now assume that the axion background field can be expanded as a power series in $\tilde{\tau}$
\begin{align}
    \bar{\phi} = \sum^{\infty}_{n=0} A_n\tilde{\tau}^n &= A_0 + A_1\tilde{\tau} + A_2\tilde{\tau}^2 \nonumber \\ &\hspace{1cm} + A_3\tilde{\tau}^3 + A_4\tilde{\tau}^4 + \mathcal{O}(\tilde{\tau}^5), \label{eq:phi_series}
\end{align}
where $A_0 = \bar{\phi}_i$ is just the initial ``misaligned'' value of the axion field. Our goal is to determine the coefficients $A_n$ order by order using perturbation theory. The time derivatives of $\bar{\phi}$ are computed as
\begin{align}
    \dfrac{\partial\bar{\phi}}{\partial\tilde{\tau}} &= A_1 + 2A_2\tilde{\tau} + 3A_3\tilde{\tau}^2 + 4A_4\tilde{\tau}^3 + \mathcal{O}(\tilde{\tau}^4), \label{eq:phi_dtau_series} \\
    \dfrac{\partial^2\bar{\phi}}{\partial\tilde{\tau}^2} &= 2A_2 + 6A_3\tilde{\tau} + 12A_4\tilde{\tau}^2 + \mathcal{O}(\tilde{\tau}^3). \label{eq:phi_ddtau_series}
\end{align}
Substituting $\bar{\phi}$ from \eqref{eq:phi_series} and its time derivatives from \eqref{eq:phi_dtau_series}, \eqref{eq:phi_ddtau_series} into Eq.~\eqref{eq:phi_ax_tau}yields an expression containing terms with various powers of $\tilde{\tau}$. We then collect terms of a particular order $\mathcal{O}(\tilde{\tau}^n)$ to construct the $n^{\rm th}$ perturbation equation as follows
\begin{align}
    &\tilde{\tau}^0-\text{order:} \quad 2A_1 = 0, \\
    &\tilde{\tau}^1-\text{order:} \quad 6A_2 = 0, \\
    &\tilde{\tau}^2-\text{order:} \quad 12A_3 = 0, \\
    &\tilde{\tau}^3-\text{order:} \quad 20A_4 + \dfrac{m^2}{\mathcal{C}^2}\bar{\phi}_i = 0.
\end{align}
These results suggest that the power-series solution of $\bar{\phi}$, truncated at the first non-vanishing term, is given by
\begin{align}
    \bar{\phi} = \bar{\phi}_i\left( 1 - \dfrac{m^2}{20\mathcal{C}^2}\tilde{\tau}^4 \right).
\end{align}
Given that $\partial\bar{\phi}/\partial\tilde{\tau} = (a/\mathcal{C})\dot{\bar{\phi}}$, it is straightforward to compute the time derivative from this solution
\begin{align}
    \dfrac{\partial\bar{\phi}}{\partial\tilde{\tau}} = - \dfrac{m^2}{5\mathcal{C}^2}\tilde{\tau}^3 \Hquad \rightarrow \Hquad \dot{\bar{\phi}} = - \dfrac{m^2\mathcal{C}^2}{5a}\bar{\phi}_i\tau^3. \label{eq:phi_dot_ic}
\end{align}

At the perturbative level, the derivation for $\delta\phi$ follows the same idea. We need to find the expression for the adiabatic growing mode of the metric perturbations, which is given by~\cite{Bucher:1999re}
\begin{align}
    h = \dfrac{1}{2}k^2\tau^2 \Hquad \rightarrow \Hquad \dfrac{\partial h}{\partial\tilde{\tau}} = \dfrac{k^2}{\mathcal{C}^2}\tilde{\tau}.
\end{align}
Plugging in $\partial\bar{\phi}/\partial\tilde{\tau}$ and $\partial h/\partial\tilde{\tau}$ as given above, the perturbation equation \eqref{eq:dphi_ax} transforms to
\begin{align}
    \tilde{\tau}\dfrac{\partial^2\delta\phi}{\partial\tilde{\tau}^2} + 2\dfrac{\partial\delta\phi}{\partial\tilde{\tau}} + \left( \dfrac{k^2}{\mathcal{C}^2}\tilde{\tau} +\dfrac{m^2}{\mathcal{C}^2}\tilde{\tau}^3 \right)\delta\phi = \dfrac{m^2k^2}{10\mathcal{C}^4}\bar{\phi}_i\tilde{\tau}^5. \label{eq:dphi_ax_tau}
\end{align}
Assuming the perturbation field can be expanded as
\begin{align}
    \delta{\phi} = \sum^{\infty}_{n=1} B_n\tilde{\tau}^n, \label{eq:dphi_series}
\end{align} 
where the $n = 0$ is omitted because the adiabatic $\delta\phi$ vanishes at zeroth order. We then substitute \eqref{eq:dphi_series} into \eqref{eq:dphi_ax_tau} to derive 
\begin{align}
    &\tilde{\tau}^0-\text{order:} \quad B_1 = 0, \\
    &\tilde{\tau}^1-\text{order:} \quad 6B_2 = 0, \\
    &\tilde{\tau}^2-\text{order:} \quad 12B_3 + \dfrac{k^2}{\mathcal{C}^2}B_1 = 0, \\
    &\tilde{\tau}^3-\text{order:} \quad 20B_4 + \dfrac{k^2}{\mathcal{C}^2}B_2 = 0, \\
    &\tilde{\tau}^4-\text{order:} \quad 30B_5 + \dfrac{k^2}{\mathcal{C}^2}B_3 + \dfrac{m^2}{\mathcal{C}^2}B_1 = 0, \\
    &\tilde{\tau}^5-\text{order:} \quad 42B_6 + \dfrac{k^2}{\mathcal{C}^2}B_4 + \dfrac{m^2}{\mathcal{C}^2}B_2 = \dfrac{m^2k^2}{10\mathcal{C}^4}\bar{\phi}_i.
\end{align}
From the above equations, we can show that $B_6$ is the first non-vanishing coefficient. As such, the power-series solution of $\delta\phi$ and its time derivative take the following forms to the leading order
\begin{align}
    &\delta\phi = \dfrac{m^2k^2}{420\mathcal{C}^4}\bar{\phi}_i\tilde{\tau}^6 = \dfrac{m^2k^2\mathcal{C}^2}{420}\bar{\phi}_i\tau^6, \label{eq:dphi_ic} \\
    &\dfrac{\partial\delta\phi}{\partial\tilde{\tau}} = \dfrac{m^2k^2}{70\mathcal{C}^4}\bar{\phi}_i\tilde{\tau}^5 \Hquad \rightarrow \Hquad \dot{\delta\phi} = \dfrac{m^2k^2\mathcal{C}^2}{70a}\bar{\phi}_i\tau^5. \label{eq:dphi_dot_ic}
\end{align}

Under normal circumstances, choosing the initial time $\tilde{\tau}_i$ sufficiently early ensures that the series expansions in \eqref{eq:phi_series} and \eqref{eq:dphi_series} converge, as $\tilde{\tau} \ll 1$ here. That said, in some cosmological codes like \texttt{CLASS}, $\tau_i$ is not chosen uniformly, as perturbations are initialized considerably later than the background by default. This becomes problematic in case $\delta\phi$ is initialized after the transition time of the background field. To see why, note that Eq.\eqref{eq:dphi_ax_tau} is derived by substituting $\partial\bar{\phi}/\partial\tilde{\tau}$ from Eq.\eqref{eq:phi_dot_ic}. However, at $\tau_i \sim \tau^b_*$, where $\tau^b_*$ denotes the background transition time, the background field has already entered the regime of damped oscillations, and the power-series solution \eqref{eq:phi_dot_ic} no longer accurately describes $\partial\bar{\phi}/\partial\tilde{\tau}$. As a result, the solutions in \eqref{eq:dphi_ic} and \eqref{eq:dphi_dot_ic} are also rendered invalid. To avoid this issue, \texttt{CLAxions} imposes a condition on perturbations of {\it all} wavenumbers to be initialized at $\tilde{\tau}_i = 0.01\tilde{\tau}^b_*$ or at the value associated with the initial scale factor $a_i = 10^{-5}$, whichever is smaller.


\bibliography{reference}

\begin{thebibliography}{30}%
\makeatletter
\providecommand \@ifxundefined [1]{%
 \@ifx{#1\undefined}
}%
\providecommand \@ifnum [1]{%
 \ifnum #1\expandafter \@firstoftwo
 \else \expandafter \@secondoftwo
 \fi
}%
\providecommand \@ifx [1]{%
 \ifx #1\expandafter \@firstoftwo
 \else \expandafter \@secondoftwo
 \fi
}%
\providecommand \natexlab [1]{#1}%
\providecommand \enquote  [1]{``#1''}%
\providecommand \bibnamefont  [1]{#1}%
\providecommand \bibfnamefont [1]{#1}%
\providecommand \citenamefont [1]{#1}%
\providecommand \href@noop [0]{\@secondoftwo}%
\providecommand \href [0]{\begingroup \@sanitize@url \@href}%
\providecommand \@href[1]{\@@startlink{#1}\@@href}%
\providecommand \@@href[1]{\endgroup#1\@@endlink}%
\providecommand \@sanitize@url [0]{\catcode `\\12\catcode `\$12\catcode
  `\&12\catcode `\#12\catcode `\^12\catcode `\_12\catcode `\%12\relax}%
\providecommand \@@startlink[1]{}%
\providecommand \@@endlink[0]{}%
\providecommand \url  [0]{\begingroup\@sanitize@url \@url }%
\providecommand \@url [1]{\endgroup\@href {#1}{\urlprefix }}%
\providecommand \urlprefix  [0]{URL }%
\providecommand \Eprint [0]{\href }%
\providecommand \doibase [0]{https://doi.org/}%
\providecommand \selectlanguage [0]{\@gobble}%
\providecommand \bibinfo  [0]{\@secondoftwo}%
\providecommand \bibfield  [0]{\@secondoftwo}%
\providecommand \translation [1]{[#1]}%
\providecommand \BibitemOpen [0]{}%
\providecommand \bibitemStop [0]{}%
\providecommand \bibitemNoStop [0]{.\EOS\space}%
\providecommand \EOS [0]{\spacefactor3000\relax}%
\providecommand \BibitemShut  [1]{\csname bibitem#1\endcsname}%
\let\auto@bib@innerbib\@empty
\bibitem [{\citenamefont {Salehian}\ \emph {et~al.}(2020)\citenamefont
  {Salehian}, \citenamefont {Namjoo},\ and\ \citenamefont
  {Kaiser}}]{Salehian:2020bon}%
  \BibitemOpen
  \bibfield  {author} {\bibinfo {author} {\bibfnamefont {B.}~\bibnamefont
  {Salehian}}, \bibinfo {author} {\bibfnamefont {M.~H.}\ \bibnamefont
  {Namjoo}},\ and\ \bibinfo {author} {\bibfnamefont {D.~I.}\ \bibnamefont
  {Kaiser}},\ }\bibfield  {title} {\bibinfo {title} {{Effective theories for a
  nonrelativistic field in an expanding universe: Induced self-interaction,
  pressure, sound speed, and viscosity}},\ }\href
  {https://doi.org/10.1007/JHEP07(2020)059} {\bibfield  {journal} {\bibinfo
  {journal} {JHEP}\ }\textbf {\bibinfo {volume} {07}},\ \bibinfo {pages}
  {059}},\ \Eprint {https://arxiv.org/abs/2005.05388} {arXiv:2005.05388
  [astro-ph.CO]} \BibitemShut {NoStop}%
\bibitem [{\citenamefont {Luu}\ and\ \citenamefont
  {Prescod-Weinstein}(2026)}]{Luu:2026las}%
  \BibitemOpen
  \bibfield  {author} {\bibinfo {author} {\bibfnamefont {H.~N.}\ \bibnamefont
  {Luu}}\ and\ \bibinfo {author} {\bibfnamefont {C.}~\bibnamefont
  {Prescod-Weinstein}},\ }\bibfield  {title} {\bibinfo {title} {{A Tale of Two
  Gauges: Effective Field Theory for Relativistic Behavior of Cosmological
  Axions}},\ }\href@noop {} {\  (\bibinfo {year} {2026})},\ \Eprint
  {https://arxiv.org/abs/2609.11055} {arXiv:2609.11055 [astro-ph.CO]}
  \BibitemShut {NoStop}%
\bibitem [{\citenamefont {Aghanim}\ \emph {et~al.}(2020)\citenamefont {Aghanim}
  \emph {et~al.}}]{Planck:2018vyg}%
  \BibitemOpen
  \bibfield  {author} {\bibinfo {author} {\bibfnamefont {N.}~\bibnamefont
  {Aghanim}} \emph {et~al.} (\bibinfo {collaboration} {Planck}),\ }\bibfield
  {title} {\bibinfo {title} {{Planck 2018 results. VI. Cosmological
  parameters}},\ }\href {https://doi.org/10.1051/0004-6361/201833910}
  {\bibfield  {journal} {\bibinfo  {journal} {Astron. Astrophys.}\ }\textbf
  {\bibinfo {volume} {641}},\ \bibinfo {pages} {A6} (\bibinfo {year} {2020})},\
  \bibinfo {note} {[Erratum: Astron.Astrophys. 652, C4 (2021)]},\ \Eprint
  {https://arxiv.org/abs/1807.06209} {arXiv:1807.06209 [astro-ph.CO]}
  \BibitemShut {NoStop}%
\bibitem [{\citenamefont {Scognamiglio}\ \emph {et~al.}(2026)\citenamefont
  {Scognamiglio} \emph {et~al.}}]{Scognamiglio:2026phv}%
  \BibitemOpen
  \bibfield  {author} {\bibinfo {author} {\bibfnamefont {D.}~\bibnamefont
  {Scognamiglio}} \emph {et~al.},\ }\bibfield  {title} {\bibinfo {title} {{An
  ultra-high-resolution map of (dark) matter}},\ }\href
  {https://doi.org/10.1038/s41550-025-02763-9} {\bibfield  {journal} {\bibinfo
  {journal} {Nature Astron.}\ }\textbf {\bibinfo {volume} {10}},\ \bibinfo
  {pages} {573} (\bibinfo {year} {2026})},\ \Eprint
  {https://arxiv.org/abs/2601.17239} {arXiv:2601.17239 [astro-ph.CO]}
  \BibitemShut {NoStop}%
\bibitem [{\citenamefont {Riess}\ \emph {et~al.}(1998)\citenamefont {Riess}
  \emph {et~al.}}]{SupernovaSearchTeam:1998fmf}%
  \BibitemOpen
  \bibfield  {author} {\bibinfo {author} {\bibfnamefont {A.~G.}\ \bibnamefont
  {Riess}} \emph {et~al.} (\bibinfo {collaboration} {Supernova Search Team}),\
  }\bibfield  {title} {\bibinfo {title} {{Observational evidence from
  supernovae for an accelerating universe and a cosmological constant}},\
  }\href {https://doi.org/10.1086/300499} {\bibfield  {journal} {\bibinfo
  {journal} {Astron. J.}\ }\textbf {\bibinfo {volume} {116}},\ \bibinfo {pages}
  {1009} (\bibinfo {year} {1998})},\ \Eprint
  {https://arxiv.org/abs/astro-ph/9805201} {arXiv:astro-ph/9805201}
  \BibitemShut {NoStop}%
\bibitem [{\citenamefont {Bradac}\ \emph {et~al.}(2008)\citenamefont {Bradac},
  \citenamefont {Allen}, \citenamefont {Treu}, \citenamefont {Ebeling},
  \citenamefont {Massey}, \citenamefont {Morris}, \citenamefont {von~der
  Linden},\ and\ \citenamefont {Applegate}}]{Bradac:2008eu}%
  \BibitemOpen
  \bibfield  {author} {\bibinfo {author} {\bibfnamefont {M.}~\bibnamefont
  {Bradac}}, \bibinfo {author} {\bibfnamefont {S.~W.}\ \bibnamefont {Allen}},
  \bibinfo {author} {\bibfnamefont {T.}~\bibnamefont {Treu}}, \bibinfo {author}
  {\bibfnamefont {H.}~\bibnamefont {Ebeling}}, \bibinfo {author} {\bibfnamefont
  {R.}~\bibnamefont {Massey}}, \bibinfo {author} {\bibfnamefont {R.~G.}\
  \bibnamefont {Morris}}, \bibinfo {author} {\bibfnamefont {A.}~\bibnamefont
  {von~der Linden}},\ and\ \bibinfo {author} {\bibfnamefont {D.}~\bibnamefont
  {Applegate}},\ }\bibfield  {title} {\bibinfo {title} {{Revealing the
  properties of dark matter in the merging cluster MACSJ0025.4-1222}},\ }\href
  {https://doi.org/10.1086/591246} {\bibfield  {journal} {\bibinfo  {journal}
  {Astrophys. J.}\ }\textbf {\bibinfo {volume} {687}},\ \bibinfo {pages} {959}
  (\bibinfo {year} {2008})},\ \Eprint {https://arxiv.org/abs/0806.2320}
  {arXiv:0806.2320 [astro-ph]} \BibitemShut {NoStop}%
\bibitem [{\citenamefont {Abdalla}\ \emph {et~al.}(2022)\citenamefont {Abdalla}
  \emph {et~al.}}]{Abdalla:2022yfr}%
  \BibitemOpen
  \bibfield  {author} {\bibinfo {author} {\bibfnamefont {E.}~\bibnamefont
  {Abdalla}} \emph {et~al.},\ }\bibfield  {title} {\bibinfo {title} {{Cosmology
  intertwined: A review of the particle physics, astrophysics, and cosmology
  associated with the cosmological tensions and anomalies}},\ }\href
  {https://doi.org/10.1016/j.jheap.2022.04.002} {\bibfield  {journal} {\bibinfo
   {journal} {JHEAp}\ }\textbf {\bibinfo {volume} {34}},\ \bibinfo {pages} {49}
  (\bibinfo {year} {2022})},\ \Eprint {https://arxiv.org/abs/2203.06142}
  {arXiv:2203.06142 [astro-ph.CO]} \BibitemShut {NoStop}%
\bibitem [{\citenamefont {Adams}\ \emph {et~al.}(2022)\citenamefont {Adams}
  \emph {et~al.}}]{Adams:2022pbo}%
  \BibitemOpen
  \bibfield  {author} {\bibinfo {author} {\bibfnamefont {C.~B.}\ \bibnamefont
  {Adams}} \emph {et~al.},\ }\bibfield  {title} {\bibinfo {title} {{Axion Dark
  Matter}},\ }in\ \href@noop {} {\emph {\bibinfo {booktitle} {{Snowmass
  2021}}}}\ (\bibinfo {year} {2022})\ \Eprint
  {https://arxiv.org/abs/2203.14923} {arXiv:2203.14923 [hep-ex]} \BibitemShut
  {NoStop}%
\bibitem [{\citenamefont {Luu}\ \emph {et~al.}(2025)\citenamefont {Luu},
  \citenamefont {Qiu},\ and\ \citenamefont {Tye}}]{Luu:2025fgw}%
  \BibitemOpen
  \bibfield  {author} {\bibinfo {author} {\bibfnamefont {H.~N.}\ \bibnamefont
  {Luu}}, \bibinfo {author} {\bibfnamefont {Y.-C.}\ \bibnamefont {Qiu}},\ and\
  \bibinfo {author} {\bibfnamefont {S.~H.~H.}\ \bibnamefont {Tye}},\ }\bibfield
   {title} {\bibinfo {title} {{Dynamical dark energy from an ultralight
  axion}},\ }\href {https://doi.org/10.1103/3mpg-24d2} {\bibfield  {journal}
  {\bibinfo  {journal} {Phys. Rev. D}\ }\textbf {\bibinfo {volume} {112}},\
  \bibinfo {pages} {023524} (\bibinfo {year} {2025})},\ \Eprint
  {https://arxiv.org/abs/2503.18120} {arXiv:2503.18120 [hep-ph]} \BibitemShut
  {NoStop}%
\bibitem [{\citenamefont {Hui}(2021)}]{Hui:2021tkt}%
  \BibitemOpen
  \bibfield  {author} {\bibinfo {author} {\bibfnamefont {L.}~\bibnamefont
  {Hui}},\ }\bibfield  {title} {\bibinfo {title} {{Wave Dark Matter}},\ }\href
  {https://doi.org/10.1146/annurev-astro-120920-010024} {\bibfield  {journal}
  {\bibinfo  {journal} {Ann. Rev. Astron. Astrophys.}\ }\textbf {\bibinfo
  {volume} {59}},\ \bibinfo {pages} {247} (\bibinfo {year} {2021})},\ \Eprint
  {https://arxiv.org/abs/2101.11735} {arXiv:2101.11735 [astro-ph.CO]}
  \BibitemShut {NoStop}%
\bibitem [{\citenamefont {Ferreira}(2021)}]{Ferreira:2020fam}%
  \BibitemOpen
  \bibfield  {author} {\bibinfo {author} {\bibfnamefont {E.~G.~M.}\
  \bibnamefont {Ferreira}},\ }\bibfield  {title} {\bibinfo {title}
  {{Ultra-light dark matter}},\ }\href
  {https://doi.org/10.1007/s00159-021-00135-6} {\bibfield  {journal} {\bibinfo
  {journal} {Astron. Astrophys. Rev.}\ }\textbf {\bibinfo {volume} {29}},\
  \bibinfo {pages} {7} (\bibinfo {year} {2021})},\ \Eprint
  {https://arxiv.org/abs/2005.03254} {arXiv:2005.03254 [astro-ph.CO]}
  \BibitemShut {NoStop}%
\bibitem [{\citenamefont {Preskill}\ \emph {et~al.}(1983)\citenamefont
  {Preskill}, \citenamefont {Wise},\ and\ \citenamefont
  {Wilczek}}]{Preskill:1982cy}%
  \BibitemOpen
  \bibfield  {author} {\bibinfo {author} {\bibfnamefont {J.}~\bibnamefont
  {Preskill}}, \bibinfo {author} {\bibfnamefont {M.~B.}\ \bibnamefont {Wise}},\
  and\ \bibinfo {author} {\bibfnamefont {F.}~\bibnamefont {Wilczek}},\
  }\bibfield  {title} {\bibinfo {title} {{Cosmology of the Invisible Axion}},\
  }\href {https://doi.org/10.1016/0370-2693(83)90637-8} {\bibfield  {journal}
  {\bibinfo  {journal} {Phys. Lett. B}\ }\textbf {\bibinfo {volume} {120}},\
  \bibinfo {pages} {127} (\bibinfo {year} {1983})}\BibitemShut {NoStop}%
\bibitem [{\citenamefont {Ratra}(1991)}]{Ratra:1990me}%
  \BibitemOpen
  \bibfield  {author} {\bibinfo {author} {\bibfnamefont {B.}~\bibnamefont
  {Ratra}},\ }\bibfield  {title} {\bibinfo {title} {{Expressions for linearized
  perturbations in a massive scalar field dominated cosmological model}},\
  }\href {https://doi.org/10.1103/PhysRevD.44.352} {\bibfield  {journal}
  {\bibinfo  {journal} {Phys. Rev. D}\ }\textbf {\bibinfo {volume} {44}},\
  \bibinfo {pages} {352} (\bibinfo {year} {1991})}\BibitemShut {NoStop}%
\bibitem [{\citenamefont {Hwang}\ and\ \citenamefont
  {Noh}(2009)}]{Hwang:2009js}%
  \BibitemOpen
  \bibfield  {author} {\bibinfo {author} {\bibfnamefont {J.-c.}\ \bibnamefont
  {Hwang}}\ and\ \bibinfo {author} {\bibfnamefont {H.}~\bibnamefont {Noh}},\
  }\bibfield  {title} {\bibinfo {title} {{Axion as a Cold Dark Matter
  candidate}},\ }\href {https://doi.org/10.1016/j.physletb.2009.08.031}
  {\bibfield  {journal} {\bibinfo  {journal} {Phys. Lett. B}\ }\textbf
  {\bibinfo {volume} {680}},\ \bibinfo {pages} {1} (\bibinfo {year} {2009})},\
  \Eprint {https://arxiv.org/abs/0902.4738} {arXiv:0902.4738 [astro-ph.CO]}
  \BibitemShut {NoStop}%
\bibitem [{\citenamefont {Hlozek}\ \emph {et~al.}(2015)\citenamefont {Hlozek},
  \citenamefont {Grin}, \citenamefont {Marsh},\ and\ \citenamefont
  {Ferreira}}]{Hlozek:2014lca}%
  \BibitemOpen
  \bibfield  {author} {\bibinfo {author} {\bibfnamefont {R.}~\bibnamefont
  {Hlozek}}, \bibinfo {author} {\bibfnamefont {D.}~\bibnamefont {Grin}},
  \bibinfo {author} {\bibfnamefont {D.~J.~E.}\ \bibnamefont {Marsh}},\ and\
  \bibinfo {author} {\bibfnamefont {P.~G.}\ \bibnamefont {Ferreira}},\
  }\bibfield  {title} {\bibinfo {title} {{A search for ultralight axions using
  precision cosmological data}},\ }\href
  {https://doi.org/10.1103/PhysRevD.91.103512} {\bibfield  {journal} {\bibinfo
  {journal} {Phys. Rev. D}\ }\textbf {\bibinfo {volume} {91}},\ \bibinfo
  {pages} {103512} (\bibinfo {year} {2015})},\ \Eprint
  {https://arxiv.org/abs/1410.2896} {arXiv:1410.2896 [astro-ph.CO]}
  \BibitemShut {NoStop}%
\bibitem [{\citenamefont {Cookmeyer}\ \emph {et~al.}(2020)\citenamefont
  {Cookmeyer}, \citenamefont {Cookmeyer}, \citenamefont {Grin},\ and\
  \citenamefont {Smith}}]{Cookmeyer:2019rna}%
  \BibitemOpen
  \bibfield  {author} {\bibinfo {author} {\bibfnamefont {T.}~\bibnamefont
  {Cookmeyer}}, \bibinfo {author} {\bibfnamefont {J.}~\bibnamefont
  {Cookmeyer}}, \bibinfo {author} {\bibfnamefont {D.}~\bibnamefont {Grin}},\
  and\ \bibinfo {author} {\bibfnamefont {T.~L.}\ \bibnamefont {Smith}},\
  }\bibfield  {title} {\bibinfo {title} {{How sound are our ultralight axion
  approximations?}},\ }\href {https://doi.org/10.1103/PhysRevD.101.023501}
  {\bibfield  {journal} {\bibinfo  {journal} {Phys. Rev. D}\ }\textbf {\bibinfo
  {volume} {101}},\ \bibinfo {pages} {023501} (\bibinfo {year} {2020})},\
  \Eprint {https://arxiv.org/abs/1909.11094} {arXiv:1909.11094 [astro-ph.CO]}
  \BibitemShut {NoStop}%
\bibitem [{\citenamefont {Passaglia}\ and\ \citenamefont
  {Hu}(2022)}]{Passaglia:2022bcr}%
  \BibitemOpen
  \bibfield  {author} {\bibinfo {author} {\bibfnamefont {S.}~\bibnamefont
  {Passaglia}}\ and\ \bibinfo {author} {\bibfnamefont {W.}~\bibnamefont {Hu}},\
  }\bibfield  {title} {\bibinfo {title} {{Accurate effective fluid
  approximation for ultralight axions}},\ }\href
  {https://doi.org/10.1103/PhysRevD.105.123529} {\bibfield  {journal} {\bibinfo
   {journal} {Phys. Rev. D}\ }\textbf {\bibinfo {volume} {105}},\ \bibinfo
  {pages} {123529} (\bibinfo {year} {2022})},\ \Eprint
  {https://arxiv.org/abs/2201.10238} {arXiv:2201.10238 [astro-ph.CO]}
  \BibitemShut {NoStop}%
\bibitem [{\citenamefont {Liu}\ \emph {et~al.}(2025)\citenamefont {Liu},
  \citenamefont {Hu},\ and\ \citenamefont {Grin}}]{Liu:2024yne}%
  \BibitemOpen
  \bibfield  {author} {\bibinfo {author} {\bibfnamefont {R.}~\bibnamefont
  {Liu}}, \bibinfo {author} {\bibfnamefont {W.}~\bibnamefont {Hu}},\ and\
  \bibinfo {author} {\bibfnamefont {D.}~\bibnamefont {Grin}},\ }\bibfield
  {title} {\bibinfo {title} {{Accurate method for ultralight axion CMB and
  matter power spectra}},\ }\href {https://doi.org/10.1103/1z4c-1w7f}
  {\bibfield  {journal} {\bibinfo  {journal} {Phys. Rev. D}\ }\textbf {\bibinfo
  {volume} {112}},\ \bibinfo {pages} {023513} (\bibinfo {year} {2025})},\
  \Eprint {https://arxiv.org/abs/2412.15192} {arXiv:2412.15192 [astro-ph.CO]}
  \BibitemShut {NoStop}%
\bibitem [{\citenamefont {Namjoo}\ \emph {et~al.}(2018)\citenamefont {Namjoo},
  \citenamefont {Guth},\ and\ \citenamefont {Kaiser}}]{Namjoo:2017nia}%
  \BibitemOpen
  \bibfield  {author} {\bibinfo {author} {\bibfnamefont {M.~H.}\ \bibnamefont
  {Namjoo}}, \bibinfo {author} {\bibfnamefont {A.~H.}\ \bibnamefont {Guth}},\
  and\ \bibinfo {author} {\bibfnamefont {D.~I.}\ \bibnamefont {Kaiser}},\
  }\bibfield  {title} {\bibinfo {title} {{Relativistic Corrections to
  Nonrelativistic Effective Field Theories}},\ }\href
  {https://doi.org/10.1103/PhysRevD.98.016011} {\bibfield  {journal} {\bibinfo
  {journal} {Phys. Rev. D}\ }\textbf {\bibinfo {volume} {98}},\ \bibinfo
  {pages} {016011} (\bibinfo {year} {2018})},\ \Eprint
  {https://arxiv.org/abs/1712.00445} {arXiv:1712.00445 [hep-ph]} \BibitemShut
  {NoStop}%
\bibitem [{\citenamefont {Salehian}\ \emph {et~al.}(2021)\citenamefont
  {Salehian}, \citenamefont {Zhang}, \citenamefont {Amin}, \citenamefont
  {Kaiser},\ and\ \citenamefont {Namjoo}}]{Salehian:2021khb}%
  \BibitemOpen
  \bibfield  {author} {\bibinfo {author} {\bibfnamefont {B.}~\bibnamefont
  {Salehian}}, \bibinfo {author} {\bibfnamefont {H.-Y.}\ \bibnamefont {Zhang}},
  \bibinfo {author} {\bibfnamefont {M.~A.}\ \bibnamefont {Amin}}, \bibinfo
  {author} {\bibfnamefont {D.~I.}\ \bibnamefont {Kaiser}},\ and\ \bibinfo
  {author} {\bibfnamefont {M.~H.}\ \bibnamefont {Namjoo}},\ }\bibfield  {title}
  {\bibinfo {title} {{Beyond Schr{\"o}dinger-Poisson: nonrelativistic effective
  field theory for scalar dark matter}},\ }\href
  {https://doi.org/10.1007/JHEP09(2021)050} {\bibfield  {journal} {\bibinfo
  {journal} {JHEP}\ }\textbf {\bibinfo {volume} {09}},\ \bibinfo {pages}
  {050}},\ \Eprint {https://arxiv.org/abs/2104.10128} {arXiv:2104.10128
  [astro-ph.CO]} \BibitemShut {NoStop}%
\bibitem [{\citenamefont {Ma}\ and\ \citenamefont
  {Bertschinger}(1995)}]{Ma:1995ey}%
  \BibitemOpen
  \bibfield  {author} {\bibinfo {author} {\bibfnamefont {C.-P.}\ \bibnamefont
  {Ma}}\ and\ \bibinfo {author} {\bibfnamefont {E.}~\bibnamefont
  {Bertschinger}},\ }\bibfield  {title} {\bibinfo {title} {{Cosmological
  perturbation theory in the synchronous and conformal Newtonian gauges}},\
  }\href {https://doi.org/10.1086/176550} {\bibfield  {journal} {\bibinfo
  {journal} {Astrophys. J.}\ }\textbf {\bibinfo {volume} {455}},\ \bibinfo
  {pages} {7} (\bibinfo {year} {1995})},\ \Eprint
  {https://arxiv.org/abs/astro-ph/9506072} {arXiv:astro-ph/9506072}
  \BibitemShut {NoStop}%
\bibitem [{\citenamefont {Howlett}\ \emph {et~al.}(2012)\citenamefont
  {Howlett}, \citenamefont {Lewis}, \citenamefont {Hall},\ and\ \citenamefont
  {Challinor}}]{Howlett:2012mh}%
  \BibitemOpen
  \bibfield  {author} {\bibinfo {author} {\bibfnamefont {C.}~\bibnamefont
  {Howlett}}, \bibinfo {author} {\bibfnamefont {A.}~\bibnamefont {Lewis}},
  \bibinfo {author} {\bibfnamefont {A.}~\bibnamefont {Hall}},\ and\ \bibinfo
  {author} {\bibfnamefont {A.}~\bibnamefont {Challinor}},\ }\bibfield  {title}
  {\bibinfo {title} {{CMB power spectrum parameter degeneracies in the era of
  precision cosmology}},\ }\href
  {https://doi.org/10.1088/1475-7516/2012/04/027} {\bibfield  {journal}
  {\bibinfo  {journal} {JCAP}\ }\textbf {\bibinfo {volume} {04}},\ \bibinfo
  {pages} {027}},\ \Eprint {https://arxiv.org/abs/1201.3654} {arXiv:1201.3654
  [astro-ph.CO]} \BibitemShut {NoStop}%
\bibitem [{\citenamefont {Blas}\ \emph {et~al.}(2011)\citenamefont {Blas},
  \citenamefont {Lesgourgues},\ and\ \citenamefont {Tram}}]{Blas:2011rf}%
  \BibitemOpen
  \bibfield  {author} {\bibinfo {author} {\bibfnamefont {D.}~\bibnamefont
  {Blas}}, \bibinfo {author} {\bibfnamefont {J.}~\bibnamefont {Lesgourgues}},\
  and\ \bibinfo {author} {\bibfnamefont {T.}~\bibnamefont {Tram}},\ }\bibfield
  {title} {\bibinfo {title} {{The Cosmic Linear Anisotropy Solving System
  (CLASS) II: Approximation schemes}},\ }\href
  {https://doi.org/10.1088/1475-7516/2011/07/034} {\bibfield  {journal}
  {\bibinfo  {journal} {JCAP}\ }\textbf {\bibinfo {volume} {07}},\ \bibinfo
  {pages} {034}},\ \Eprint {https://arxiv.org/abs/1104.2933} {arXiv:1104.2933
  [astro-ph.CO]} \BibitemShut {NoStop}%
\bibitem [{\citenamefont {Marsh}(2016)}]{Marsh:2015xka}%
  \BibitemOpen
  \bibfield  {author} {\bibinfo {author} {\bibfnamefont {D.~J.~E.}\
  \bibnamefont {Marsh}},\ }\bibfield  {title} {\bibinfo {title} {{Axion
  Cosmology}},\ }\href {https://doi.org/10.1016/j.physrep.2016.06.005}
  {\bibfield  {journal} {\bibinfo  {journal} {Phys. Rept.}\ }\textbf {\bibinfo
  {volume} {643}},\ \bibinfo {pages} {1} (\bibinfo {year} {2016})},\ \Eprint
  {https://arxiv.org/abs/1510.07633} {arXiv:1510.07633 [astro-ph.CO]}
  \BibitemShut {NoStop}%
\bibitem [{\citenamefont {Khmelnitsky}\ and\ \citenamefont
  {Rubakov}(2014)}]{Khmelnitsky:2013lxt}%
  \BibitemOpen
  \bibfield  {author} {\bibinfo {author} {\bibfnamefont {A.}~\bibnamefont
  {Khmelnitsky}}\ and\ \bibinfo {author} {\bibfnamefont {V.}~\bibnamefont
  {Rubakov}},\ }\bibfield  {title} {\bibinfo {title} {{Pulsar timing signal
  from ultralight scalar dark matter}},\ }\href
  {https://doi.org/10.1088/1475-7516/2014/02/019} {\bibfield  {journal}
  {\bibinfo  {journal} {JCAP}\ }\textbf {\bibinfo {volume} {02}},\ \bibinfo
  {pages} {019}},\ \Eprint {https://arxiv.org/abs/1309.5888} {arXiv:1309.5888
  [astro-ph.CO]} \BibitemShut {NoStop}%
\bibitem [{\citenamefont {Modirzadeh}\ \emph {et~al.}(2025)\citenamefont
  {Modirzadeh}, \citenamefont {Moti},\ and\ \citenamefont
  {Namjoo}}]{Modirzadeh:2025gjd}%
  \BibitemOpen
  \bibfield  {author} {\bibinfo {author} {\bibfnamefont {H.~S.}\ \bibnamefont
  {Modirzadeh}}, \bibinfo {author} {\bibfnamefont {R.}~\bibnamefont {Moti}},\
  and\ \bibinfo {author} {\bibfnamefont {M.~H.}\ \bibnamefont {Namjoo}},\
  }\bibfield  {title} {\bibinfo {title} {{Non-relativistic effective theories
  for fields with general potentials and their implications for cosmology}}\
  }\href {https://doi.org/10.1088/1475-7516/2026/03/076}
  {10.1088/1475-7516/2026/03/076} (\bibinfo {year} {2025}),\ \Eprint
  {https://arxiv.org/abs/2507.08786} {arXiv:2507.08786 [astro-ph.CO]}
  \BibitemShut {NoStop}%
\bibitem [{\citenamefont {Hu}(2003)}]{Hu:2003hjx}%
  \BibitemOpen
  \bibfield  {author} {\bibinfo {author} {\bibfnamefont {W.}~\bibnamefont
  {Hu}},\ }\bibfield  {title} {\bibinfo {title} {{Covariant linear perturbation
  formalism}},\ }\href@noop {} {\bibfield  {journal} {\bibinfo  {journal} {ICTP
  Lect. Notes Ser.}\ }\textbf {\bibinfo {volume} {14}},\ \bibinfo {pages} {145}
  (\bibinfo {year} {2003})},\ \Eprint {https://arxiv.org/abs/astro-ph/0402060}
  {arXiv:astro-ph/0402060} \BibitemShut {NoStop}%
\bibitem [{\citenamefont {Poulin}\ \emph {et~al.}(2018)\citenamefont {Poulin},
  \citenamefont {Smith}, \citenamefont {Grin}, \citenamefont {Karwal},\ and\
  \citenamefont {Kamionkowski}}]{Poulin:2018dzj}%
  \BibitemOpen
  \bibfield  {author} {\bibinfo {author} {\bibfnamefont {V.}~\bibnamefont
  {Poulin}}, \bibinfo {author} {\bibfnamefont {T.~L.}\ \bibnamefont {Smith}},
  \bibinfo {author} {\bibfnamefont {D.}~\bibnamefont {Grin}}, \bibinfo {author}
  {\bibfnamefont {T.}~\bibnamefont {Karwal}},\ and\ \bibinfo {author}
  {\bibfnamefont {M.}~\bibnamefont {Kamionkowski}},\ }\bibfield  {title}
  {\bibinfo {title} {{Cosmological implications of ultralight axionlike
  fields}},\ }\href {https://doi.org/10.1103/PhysRevD.98.083525} {\bibfield
  {journal} {\bibinfo  {journal} {Phys. Rev. D}\ }\textbf {\bibinfo {volume}
  {98}},\ \bibinfo {pages} {083525} (\bibinfo {year} {2018})},\ \Eprint
  {https://arxiv.org/abs/1806.10608} {arXiv:1806.10608 [astro-ph.CO]}
  \BibitemShut {NoStop}%
\bibitem [{\citenamefont {Smith}\ \emph {et~al.}(2020)\citenamefont {Smith},
  \citenamefont {Poulin},\ and\ \citenamefont {Amin}}]{Smith:2019ihp}%
  \BibitemOpen
  \bibfield  {author} {\bibinfo {author} {\bibfnamefont {T.~L.}\ \bibnamefont
  {Smith}}, \bibinfo {author} {\bibfnamefont {V.}~\bibnamefont {Poulin}},\ and\
  \bibinfo {author} {\bibfnamefont {M.~A.}\ \bibnamefont {Amin}},\ }\bibfield
  {title} {\bibinfo {title} {{Oscillating scalar fields and the Hubble tension:
  a resolution with novel signatures}},\ }\href
  {https://doi.org/10.1103/PhysRevD.101.063523} {\bibfield  {journal} {\bibinfo
   {journal} {Phys. Rev. D}\ }\textbf {\bibinfo {volume} {101}},\ \bibinfo
  {pages} {063523} (\bibinfo {year} {2020})},\ \Eprint
  {https://arxiv.org/abs/1908.06995} {arXiv:1908.06995 [astro-ph.CO]}
  \BibitemShut {NoStop}%
\bibitem [{\citenamefont {Bucher}\ \emph {et~al.}(2000)\citenamefont {Bucher},
  \citenamefont {Moodley},\ and\ \citenamefont {Turok}}]{Bucher:1999re}%
  \BibitemOpen
  \bibfield  {author} {\bibinfo {author} {\bibfnamefont {M.}~\bibnamefont
  {Bucher}}, \bibinfo {author} {\bibfnamefont {K.}~\bibnamefont {Moodley}},\
  and\ \bibinfo {author} {\bibfnamefont {N.}~\bibnamefont {Turok}},\ }\bibfield
   {title} {\bibinfo {title} {{The General primordial cosmic perturbation}},\
  }\href {https://doi.org/10.1103/PhysRevD.62.083508} {\bibfield  {journal}
  {\bibinfo  {journal} {Phys. Rev. D}\ }\textbf {\bibinfo {volume} {62}},\
  \bibinfo {pages} {083508} (\bibinfo {year} {2000})},\ \Eprint
  {https://arxiv.org/abs/astro-ph/9904231} {arXiv:astro-ph/9904231}
  \BibitemShut {NoStop}%
\end{thebibliography}%

\end{document}